\documentclass[a4paper,fleqn,10pt]{article}
\pdfoutput=1
\usepackage{amsmath}
\usepackage{amssymb}
\usepackage{newtxtext,newtxmath}
\usepackage{bbold}
\usepackage{array}
\usepackage{calc}
\usepackage{longtable}
\usepackage{booktabs}
\usepackage{multirow}
\usepackage[T1]{fontenc}
\usepackage{xcolor}
\usepackage{graphicx}
\usepackage{xspace}
\usepackage{units}

\numberwithin{equation}{section}
\usepackage[pdfborder={0 0 0}]{hyperref}
\usepackage[format=hang,labelfont=bf,hypcap=true]{caption}
\usepackage{subcaption}
\usepackage{sectsty}
\usepackage{enumitem}
\usepackage{mciteplus}
\usepackage{preprint/authblk} 
\allsectionsfont{\sffamily}
\subsubsectionfont{\mdseries\itshape\large}
\makeatletter
\DeclareRobustCommand*{\bfseries}{%
  \not@math@alphabet\bfseries\mathbf
  \fontseries\bfdefault\selectfont
  \boldmath
}
\makeatother
\let\spreprint\empty
\newcommand{\preprint}[1]{\def\spreprint{\protect#1}}
\let\sinstitute\empty

\makeatletter
\renewcommand{\maketitle}{\begingroup
  \null\thispagestyle{empty}%
    \ifx\spreprint\empty
      \vskip 5ex
    \else
      \flushright\large\spreprint\vskip 10ex
    \fi
    \vskip 5ex
    \flushleft
      {\sffamily\bfseries\huge\@title}\vskip 6ex
      \@author\vskip 2ex
      \ifx\sinstitute\empty
      \else
        {\small\sinstitute}
      \fi
    \vskip 5ex
  \endgroup
}
\makeatother
\renewenvironment{abstract}{\begin{center}
  {\large\sffamily\bfseries Abstract: }
  \begin{minipage}[t]{0.75\textwidth}
}{\end{minipage}\end{center}\vskip 10ex}

\numberwithin{equation}{section}
\allowdisplaybreaks[2]
\newcommand{\eg}{\textit{e.g.\@}\xspace}
\newcommand{\ie}{\textit{i.e.\@}\xspace}

\newcommand{\MCatNLO}{M\protect\scalebox{0.8}{C}@N\protect\scalebox{0.8}{LO}\xspace}
\newcommand{\MCatNLOsud}{M\protect\scalebox{0.8}{C}@N\protect\scalebox{0.8}{LO}$^\text{sud}$\xspace}

\newcommand{\Powheg}{P\protect\scalebox{0.8}{OWHEG}\xspace}

\newcommand{\MEPSatNLO}{M\scalebox{0.8}{E}P\scalebox{0.8}{S}@N\scalebox{0.8}{LO}\xspace}

\newcommand{\KrkNLO}{K\protect\scalebox{0.8}{RK}N\protect\scalebox{0.8}{LO}\xspace}

\newcommand{\Pythia}{P\protect\scalebox{0.8}{YTHIA}\xspace}

\newcommand{\YFS}{Y\protect\scalebox{0.8}{FS}\xspace}

\newcommand{\LHAPDF}{L\protect\scalebox{0.8}{HAPDF}\xspace}

\newcommand{\MSbar}{\ensuremath{\overline{\text{MS}}}\xspace}
\newcommand{\QCD}{Q\protect\scalebox{0.8}{CD}\xspace}
\newcommand{\QED}{Q\protect\scalebox{0.8}{ED}\xspace}
\newcommand{\EW}{E\protect\scalebox{0.8}{W}\xspace}
\newcommand{\QCDpQED}{Q\protect\scalebox{0.8}{CD}+Q\protect\scalebox{0.8}{ED}\xspace}
\newcommand{\QCDpEW}{Q\protect\scalebox{0.8}{CD}+E\protect\scalebox{0.8}{W}\xspace}
\newcommand{\QCDpYFS}{Q\protect\scalebox{0.8}{CD}+Y\protect\scalebox{0.8}{FS}\xspace}

\newcommand{\CutTools}{C\protect\scalebox{0.8}{UT}T\protect\scalebox{0.8}{OOLS}\xspace}
\newcommand{\OneLoop}{O\protect\scalebox{0.8}{NE}L\protect\scalebox{0.8}{OOP}\xspace}

\newcommand{\OpenLoops}{O\protect\scalebox{0.8}{PEN}L\protect\scalebox{0.8}{OOPS}\xspace}
\newcommand{\Collier}{C\protect\scalebox{0.8}{OLLIER}\xspace}
\newcommand{\FastJet}{F\protect\scalebox{0.8}{AST}J\protect\scalebox{0.8}{ET}\xspace}
\newcommand{\Sherpa}{S\protect\scalebox{0.8}{HERPA}\xspace}
\newcommand{\Comix}{C\protect\scalebox{0.8}{OMIX}\xspace}

\newcommand{\Amegic}{A\protect\scalebox{0.8}{MEGIC}\xspace}
\newcommand{\CSShower}{C\protect\scalebox{0.8}{SSHOWER}\xspace}
\newcommand{\Alaric}{A\protect\scalebox{0.8}{LARIC}\xspace}

\long\def\symbolfootnote[#1]#2{\begingroup%
\def\thefootnote{\fnsymbol{footnote}}\footnote[#1]{#2}\endgroup}

\newcommand{\im}{\imath}
\newcommand{\jm}{\jmath}
\makeatletter
\providecommand*{\diff}%
{\@ifnextchar^{\DIfF}{\DIfF^{}}}
\def\DIfF^#1{%
\mathop{\mathrm{\mathstrut d}}%
\nolimits^{#1}\gobblespace}
\def\gobblespace{%
\futurelet\diffarg\opspace}
\def\opspace{%
\let\DiffSpace\!%
\ifx\diffarg(%
\let\DiffSpace\relax
\else
\ifx\diffarg[%
\let\DiffSpace\relax
\else
\ifx\diffarg\{%
\let\DiffSpace\relax
\fi\fi\fi\DiffSpace}
\makeatother
\newcommand{\mc}[1]{\mathcal{#1}}
\newcommand{\mr}[1]{\mathrm{#1}}
\newcommand{\mb}[1]{\mathbb{#1}}
\newcommand{\mf}[1]{\mathbf{#1}}

\newcommand{\tres}{\ensuremath{t_\text{res}}}
\newcommand{\Deltares}{\ensuremath{\Delta_\text{res}}}
\newcommand{\sinweff}{\ensuremath{\sin\theta_\text{w}^\text{eff}}}
\newcommand{\Gmu}{\ensuremath{G_\mu}}
\newcommand{\GF}{\ensuremath{G_F}}
\newcommand{\ijt}{{\widetilde{\im\hspace*{-1pt}\jm}}}

\newcommand{\kt}{{\tilde{k}}}

\newcommand{\done}{{\rm d}}
\newcommand{\order}{\mathcal{O}}
\newcommand{\alphaS}{\alpha_s}

\newcommand{\bea}{\begin{eqnarray}}
\newcommand{\eea}{\end{eqnarray}}
\newcommand{\bi}{\begin{itemize}}
\newcommand{\ei}{\end{itemize}}

\newcommand{\mhl}{\vphantom{\int_A^B}}

\newcommand*{\TeV}{\ensuremath{\text{Te\kern -0.1em V}}}
\newcommand*{\GeV}{\ensuremath{\text{Ge\kern -0.1em V}}}
\newcommand*{\MeV}{\ensuremath{\text{Me\kern -0.1em V}}}
\newcommand*{\keV}{\ensuremath{\text{ke\kern -0.1em V}}}
\newcommand*{\eV}{\ensuremath{\text{e\kern -0.1em V}}}

\newcommand{\kTsq}{\ensuremath{{\mathrm{k}_\mathrm{T}^2}}\xspace}

\newlist{myitemize}{itemize}{3}
\setlist[myitemize]{leftmargin=14em}

\newcolumntype{C}{>{\centering\arraybackslash}p{0.14\textwidth}}

\newlength{\unitcharwidth}
\hypersetup{
  pdfauthor={Lois Flower, Joanne Roper, Marek Schonherr},
  pdftitle={A resonance-aware MC@NLO QCD+EW-matched calculation of lepton-pair production}
}
\preprint{IPPP/26/55\\MCnet-26-21}
\author[1]{Lois~Flower}
\author[2]{Joanne Roper}
\author[2]{Marek~Sch\"onherr}
\affil[1]{Department of Mathematical Sciences, University of Liverpool, Liverpool L69 3BX, UK}
\affil[2]{Institute for Particle Physics Phenomenology, Department of Physics, Durham University, Durham, DH1 3LE, UK}
\title{A resonance-aware \MCatNLO \QCD{}+\EW-matched\\[3mm]calculation of lepton-pair production}
\begin{document}
\vspace*{10mm}
\maketitle
\vspace*{20mm}
\begin{abstract}
  As we approach HL-LHC, there is a growing need for increased
  precision in theoretical predictions so that meaningful
  comparisons with experimental data can be made.
  It is no longer sufficient to include only \QCD higher-order
  corrections, with \EW effects becoming increasingly important.
  Even at hadron colliders, \QED radiation provides large corrections
  to some observables.
  In this paper, we present the first automated matching of
  NLO \QCDpEW to an interleaved \QCDpQED parton shower
  using the \MCatNLO matching method in the Catani-Seymour dipole formalism.
  When considering such a matched parton shower, the presence of
  resonances can lead to spurious higher order terms, originating
  in the recoil assignment, within the standard dipole construction.
  We therefore develop a resonance-aware modification to the \MCatNLO
  algorithm that can be applied to \QCD- and \QED-singlet resonances.
  We validate our interleaved matching and its resonance-aware 
  modification against fixed-order NLO \QCDpEW and
  pure \MCatNLO \QCD combined with \YFS resummation.
  Finally, we present resonance-aware \MCatNLO \QCDpEW predictions
  for Drell-Yan lepton pair production, a vital precision process
  at hadron colliders.
\end{abstract}
\newpage
\tableofcontents
\vspace*{5mm}
\hrule
\section{Introduction}
\label{sec:intro}

We are now operating in the precision era of collider physics,
with the experimental uncertainty on many measurements now
smaller than the theoretical uncertainty associated with
Standard Model predictions.
As work begins on the HL-LHC upgrade,
and with many proposals for precision- and energy-frontier
experiments on the horizon, the uncertainty on
experimental measurements will only continue to decrease.
Thus, the theory and phenomenology community
must work to improve theoretical predictions.
In particular, Drell-Yan lepton pair production has a particularly
clean experimental signature, leading to its widespread use in
calibration, normalisation, and PDF extraction. Many experimental
quantities are therefore dependent on a precise theoretical
description of this process.

Predictions at next-to-leading order (NLO) accuracy in
Quantum Chromodynamics (\QCD) are now standard, and automated,
for processes involving coloured particles, with next-to-next-to leading
order (NNLO) \QCD calculations becoming increasingly available.
By a simple power counting, it is clear that NLO corrections
originating in the electroweak (\EW) sector of the Standard Model are
comparable to the NNLO \QCD ones and, for observables involving
colour neutral particles and or large weak isospin, sometimes far larger.

Along with fixed order calculations, we must account for the
resummation of soft and collinear logarithms, which is commonly
performed numerically using parton showers.
Various algorithms exist to automate the matching of fixed-order
NLO \QCD calculations to parton showers in a process-independent way.
These include \Powheg \cite{Nason:2004rx} and
\MCatNLO \cite{Frixione:2002ik}, as well as 
\KrkNLO \cite{Jadach:2015mza}, 
multiplicative-accumulative matching \cite{Nason:2021xke},
and ESME \cite{vanBeekveld:2025lpz}.

The \Powheg framework has been used to match NLO \QCDpEW to
a \QCD parton shower for neutral-current Drell-Yan production
\cite{Bernaciak:2012hj}, the process under consideration in this paper,
and to non-interleaved \QCD and \QED showers for both neutral-
and charged-current Drell-Yan \cite{Barze:2013fru,Barze:2012tt}.
It has also been used to match NLO \QCDpEW to 
an interleaved \QCDpQED parton shower for 
Drell-Yan \cite{Muck:2016pko},
vector boson pair production \cite{Chiesa:2020ttl},
and Higgs production with an associated lepton pair 
\cite{Granata:2017iod}.

In this paper, we extend recent work to automate the \MCatNLO method for
\EW matching \cite{Flower:2026byh} to \QCDpEW NLO+PS at hadron colliders.
We build upon the existing \MCatNLO implementation \cite{Hoeche:2011fd}
using the Catani-Seymour dipole formalism \cite{Catani:1996vz,Catani:2002hc}
within the \Sherpa event generation framework
\cite{Gleisberg:2003xi,Gleisberg:2008ta,Bothmann:2019yzt,Sherpa:2024mfk}.
In particular, we describe a modification to the \MCatNLO algorithm for
processes that contain colour- and charge-neutral resonances.

Processes that contain resonances present an additional challenge
to parton shower and matching techniques, since these rely on
recoil assignment to spectator partons. The choice of recoil assignment
affects the virtuality of the resonant propagator and can introduce
spurious higher-order, but potentially numerically large, contributions.
As such, there have been efforts to develop resonance-aware
subtraction and matching schemes, with particular emphasis on
top-quark production and a \QCD parton shower evolution
\cite{Jezo:2015aia,Frederix:2016rdc,Jezo:2023rht},
$W/Z$ resonances at NLO \QCDpEW \cite{Muck:2016pko}
and $HV$ at NLO \QCDpEW \cite{Granata:2017iod}, all
based on the FKS formalism \cite{Frixione:1995ms}.
There also exist resonance-aware schemes using
Catani-Seymour dipoles \cite{Catani:1996vz,Catani:2002hc}
for top-quark pair production \cite{Hoche:2018ouj,Denner:2026ztd} at NLO \QCD.
A width-aware \QCD NLO+PS matching scheme for $t\bar{t}$ at lepton colliders
has also been recently implemented in the NLL-accurate parton shower \Alaric \cite{Hoche:2026dup}.

This paper is structured as follows: in Sec.\ \ref{sec:methods}
we outline the interleaved resonance-aware \MCatNLO \QCDpEW matching,
beginning with the interleaved dipole shower in Sec.\ \ref{sec:methods:ps}
then introducing the algorithm to match to an NLO calculation
in Sec.\ \ref{sec:methods:mcatnlo}.
We describe the resonance-aware modifications to this
algorithm in Sec.\ \ref{sec:methods:resaware}.
Validation of these methods is shown in Sec.\ \ref{sec:results:validation},
followed by phenomenological results for $\mr{pp}\to e^+e^-$
in Sec.\ \ref{sec:results:prediction}.
Finally, Sec.\ \ref{sec:conclusions} offers some concluding remarks.

\section{Interleaved \texorpdfstring{\MCatNLO}{MC@NLO} matching}
\label{sec:methods}

We begin this section by reviewing the
interleaved \QCD and \QED dipole shower, which provides approximate
higher-order corrections by resumming the dominant logarithms
that arise in the soft-collinear limit of particle radiation.
These approximate corrections will be matched to the
exact NLO \QCD and \EW expressions using an extension of
the \MCatNLO method, as described in section \ref{sec:methods:mcatnlo}.
Whilst these methods are generally process-independent,
care must be taken when internal resonances are present
in the process under consideration.
The soft-emission eikonals which span internal resonances,
and the resulting momentum reassignments inherent
in all parton shower formulations which include these eikonals,
should not introduce uncontrolled higher order contributions to
the shape of the intermediate resonance's virtuality distribution.

\subsection{Interleaved \texorpdfstring{\QCD}{\text{QCD}} and \texorpdfstring{\QED}{\text{QED}} dipole shower evolution}
\label{sec:methods:ps}

In high energy scattering processes, soft and collinear radiation
emitted from (colour-)charged external legs is logarithmically
enhanced.
Corrections to observables which are sensitive to additional radiation
can therefore become large, especially for \QCD effects,
but \QED effects are also appreciable in certain situations.
We therefore use parton showers to resum these logarithms and,
through iteratively generating emissions in the soft and collinear
approximations, the shower provides an all-orders description of the
logarithmically enhanced phase space regions.
This allows us to describe the leading structures of additional
parton emissions much more simply than full higher
order calculations could, making them useful for realistic
experimental simulations.
For convenience, we use the term \emph{partons} to refer to particles
that partake in short-distance \QED and \QCD interactions.\footnote{%
  Short-distance partons and measurable long-distance particles
  are connected through parton distribution functions in the initial state and fragmentation functions in the final state.
  While this delineation is natural in \QCD calculation due to
  confinement, with quarks and gluons denoting the short-distance
  (partonic) degrees of freedom and pions, kaons, protons, etc,
  the long-distance ones, particular care has to be taken in the
  electroweak sector where conventional nomenclature uses the same
  names -- electrons, muons, photons -- for both short-distance (partonic)
  and long-distance fields.
}

The parton shower evolution is defined recursively through a generating
functional, $\mc{F}_n(t_n;O)$ where $t$ is the ordering parameter
of the shower and $O$ is some IR-safe observable defined on the
$n$-parton phase space $\Phi_n$.
The choice of $t$ in terms of parton momenta must reflect the
hardness of the emission, \ie be able to correctly identify the
relevant soft and collinear limits, such that decreasing evolution
in $t$ leads to the production of softer and/or more collinear
radiation.
Here, we choose $t$ to be the transverse momentum generated in the
emission of
massless gauge bosons, and to be the virtuality of the splitting
parton when gauge bosons split
into fermion-anti-fermion pairs.
Following the notation of \cite{Hoeche:2011fd}, the expectation
value of $O$ is then given by
\begin{equation}\label{eq:methods:ps:lops_obs}
  \langle O \rangle^{\text{LOPS}}
  =
    \int\done\Phi_n\;\mr{B}_n(\Phi_n)\otimes\mc{F}_n(t_n;O)\,,
\end{equation}
a convolution (as indicated by the $\otimes$ symbol in
the convention of \cite{Catani:1996vz,Catani:2002hc})
of the Born matrix element $\mr{B}$ (including all PDF
and flux factors) and the
generating functional of the parton shower.
The latter is given by
\begin{equation}\label{eq:methods:ps:psfunc}
  \mc{F}_n(t_n;O)
  =
    \Delta_n(t_n,t_c)\,O(\Phi_n)
    +\int_{t_c}^{t_n}
     \done\Phi_1\,
     \mr{K}_n(\Phi_1)\,
     \Delta_n(t_n,t_{n+1})\,
     \mc{F}_{n+1}(t_{n+1};O)\;.
\end{equation}
Therein, $\done\Phi_1=\done\Phi_1(t,z,\phi)=
\frac{\done t}{t}\done z\frac{\done\phi}{2\pi}\,J(t,z,\phi)$
is the one-emission phase space element, where the evolution
scale $t$ is the ordering parameter described above, $z$ is the
splitting variable describing how momentum is shared between
the daughter partons, $\phi$
is the azimuthal angle generated in the splitting process,
and $J$ is the Jacobian
of the factorised $(n\!+\!1)$-parton phase space,
$\Phi_{n+1}=\Phi_n\!\cdot\Phi_1$.
As the relevant scale of the $n$-parton process, $t_n$ acts
as the parton shower starting scale, while
$t_{n+1}=t(\Phi_{n+1})$ is the scale associated with
the newly generated emission
and $t_c$ is an infrared cut-off below which any emission is
considered unresolvable. The inclusion of a cut-off scale renders
all integrals finite in four spacetime dimensions.
This functional describes the splitting process from the
$n$-parton state at scale
$t_{n+1}$ with the help of the splitting kernels,
\begin{equation}\label{eq:methods:ps:pskernels}
  \mr{K}_n(\Phi_1)
  =
    \sum_{a\in\{\text{\QCD,\QED}\}}
    \mr{K}_n^a(\Phi_1)\,,
\end{equation}
encoding both \QCD and \QED splittings.
In consequence, $\mr{K}_n^\text{\QCD}$ is of $\order(\alphaS)$
while $\mr{K}_n^\text{\QED}$ is of $\order(\alpha)$.
Conversely, the Sudakov form factor of the $n$-parton
configuration,
\begin{equation}\label{eq:methods:ps:pssudakov}
  \Delta_n(t,t')
  =
    \exp\left[
      -\int_{t'}^{t}\done\Phi_1\,\mr{K}_n(\Phi_1)
    \right]
  =
    \exp\left[
      -\int_{t'}^{t}\done\Phi_1\,
      \!\!\!\!\sum_{a\in\{\text{\QCD,\QED}\}}\!\!\!\!
      \mr{K}_n^a(\Phi_1)
    \right]
  =
    \prod_{a\in\{\text{\QCD,\QED}\}}\!\!\!\!
    \Delta_n^a(t,t')
    \,,
\end{equation}
describes the probability of no further splitting occurring
between scales $t$ and $t'$.
The Sudakov form factor resums the associated logarithms
arising in the soft-collinear limit.
It is the product of the individual \QCD and \QED Sudakov
form factors, $\Delta_n^\text{\QCD}$ and $\Delta_n^\text{\QED}$.
The parton shower generating functional $\mc{F}$ therefore
cannot be decomposed into independent \QCD and \QED
showers, but forms an interleaved \QCDpQED evolution.

\begin{figure}[t!]
\centering
  \begin{subfigure}[b]{0.45\linewidth}
  \includegraphics{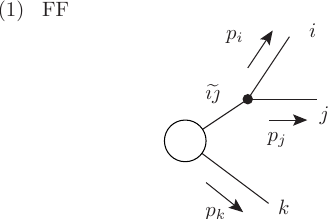}
  \end{subfigure}
  \begin{subfigure}[b]{0.45\linewidth}
  \includegraphics{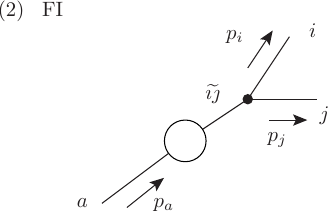}
  \end{subfigure}

  \begin{subfigure}[b]{0.45\linewidth}
  \includegraphics{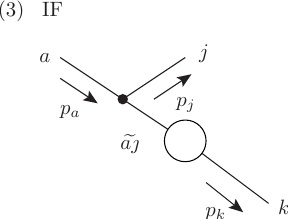}
  \end{subfigure}
  \begin{subfigure}[b]{0.45\linewidth}
  \includegraphics{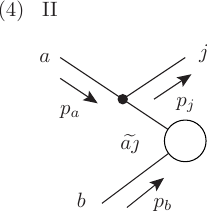}
  \end{subfigure}
  \caption{
    Diagrammatic representation of all four dipole types.
    \label{fig:methods:ps:dips}
  }
\end{figure}

We implement this evolution as an interleaved dipole shower,
extending the existing \CSShower framework of \cite{Flower:2026byh,Schumann:2007mg,Hoeche:2009xc}
based on Catani-Seymour dipoles \cite{Catani:1996vz,Catani:2002hc,Schonherr:2017qcj}.
For the dipole splitting process $\ijt+\kt\rightarrow i+j+k$,
wherein the emitter $\ijt$ becomes $i$ by emitting $j$ while
$\kt\rightarrow k$ is the spectator absorbing any
recoil in order to keep all partons on their mass shell at
every step of the parton shower, the dipole splitting functions
take the general form
\begin{equation}\label{eq:methods:ps:pskernels_qcd_qed}
  \mr{K}_n^a(\Phi_1)
  =
    \sum_{\{i,j,k\}} \mr{K}_{ij,k}^a(\Phi_1)
  =
    \sum_{\{i,j,k\}}
    -\frac{c_{ij,k}^a}{2p_ip_j}\,
    \mf{C}_{ij,k}^a\,\mf{V}_{ij,k}(\Phi_1)
  \,,
\end{equation}
where the sum runs over all possible pairs $\{\ijt,\kt\}$ of the
$n$-parton state as well as all possible products $\{i,j\}$
for each emitter $\ijt$.
$p_i$ and $p_j$ are the momenta of partons $i$ and $j$ respectively,
with $1/(2p_ip_j)$ therefore the divergent propagator,
and $c_{ij,k}^a$ is a coupling factor which is discussed in more detail
below.
Further, $\mf{C}_{ij,k}^a$ is the colour/charge correlator, which
is matrix-valued in colour space, and $\mf{V}_{ij,k}$ is the
helicity/polarisa\-tion-dependent splitting function, which is also
matrix-valued in helicity/polarisation space.
To use these splitting functions in a probabilistic interpretation
of our parton shower evolution, they need
to be cast into a scalar positive-definite form.
To achieve this, the spin-dependent splitting
functions $\mf{V}_{ij,k}$ are replaced by the spin-averaged
splitting functions
$\langle \mr{V}_{ij,k} \rangle$, which are given in
\cite{Flower:2026byh,Schumann:2007mg,Hoeche:2009xc}.%
\footnote{
  Please note that at variance with \cite{Catani:1996vz,Catani:2002hc,Schumann:2007mg,
    Hoeche:2009xc,Schonherr:2017qcj} we have removed
  all relevant coupling factors, $8\pi\,\alphaS$ and $8\pi\,\alpha$ as well
  as $C_F$, $C_A$, $T_R$, etc, from the definition of $\mf{V}$
  here and put them in the separate coupling factor $c_{ij,k}$.
  Drawing a distinction between emitter $i$ and emittee $j$,
  \ie delineating $g^*\to g_1 g_2$ and $g^*\to g_2 g_1$,
  a further factor of 2 has been removed for $g\to gg$ splittings
  in order separate its two collinear limits.
}
The \QCD colour correlator is customarily cast into scalar form
by turning
to the large-$N_c$ limit, in which \cite{Schumann:2007mg}
\begin{equation}
  \label{eq:methods:ps:Cijk}
  \mf{C}_{ij,k}^\text{\QCD}
    =
      \frac{\mf{T}_{\kt}\mf{T}_{\ijt}}{\mf{T}_\ijt^2}
    =
      \left.-\frac{\mb{1}}{n_\text{spec}}
      \right|_{\ijt,\kt \text{ colour connected}}
      +O\left(\frac{1}{N_c^2}\right)
\end{equation}
where $\mf{T}_{\ijt}$ and $\mf{T}_{\kt}$ are the colour matrices
of the emitter and spectator, respectively,
and $n_\text{spec}=1(2)$ for gluon emissions (splittings)
is the number of colour connected spectators.
$\mf{T}_{\ijt}^2$ is the quadratic Casimir of the appropriate
representation and thus equal to $C_F$ for quarks,
$C_A$ for gluons, and zero otherwise.
Here we have a natural choice of spectator(s) in the parton(s)
colour-connected to the emitter, since the contributions from
all other possible spectators are suppressed by powers of
$\frac{1}{N_c^2}$, which can be neglected to first approximation.

For QED splittings, we instead have a scalar charge correlator,
albeit one of variable sign,
\cite{Schonherr:2017qcj,Yennie:1961ad,Dittmaier:1999mb,Dittmaier:2008md,
  Kallweit:2017khh}
\begin{equation}
  \label{eq:methods:ps:Qijk}
  \mf{C}_{ij,k}^\text{\QED}=Q_{\ijt,\kt}^2=
  \left\{
  \begin{array}{ll}
    \frac{Q_{\kt}\theta_{\kt} Q_{\ijt}\theta_{\ijt}}{Q_\ijt^2}
    & \ijt\neq\gamma \\
    \kappa_{\ijt,\kt}
    & \ijt=\gamma
  \end{array}\right.
  \qquad\qquad \sum_{\kt\neq\ijt}\kappa_{\ijt,\kt}=-1 \quad\forall\,\ijt=\gamma
\end{equation}
where $Q_{\ijt}$ and $Q_{\kt}$ are the charges of the emitter
and spectator, and $\theta=(-)1$ if they are in the final
(initial) state.
No equivalent of the leading colour approximation exists in
\QED; we must therefore account for all spectators 
to photon emissions, of which
there could be many, especially at the high parton multiplicities
in the later stages of the parton shower evolution,
with every photon splitting and some gluon splittings adding another
charged-particle pair.
The resulting occurrence of dipoles with an overall negative
splitting function can be incorporated using the weighted
shower approach of
\cite{Hoeche:2011fd,Hoeche:2009xc,Platzer:2011dq,Lonnblad:2012hz},
but leads to an increase in the fraction of negative-weighted events,
with a detrimental impact on the statistical power of the sample.

In particular, a photon or gluon of a configuration with
$n_+$ positively-charged and $n_-$ negatively-charged partons
which splits into a fermion-antifermion pair will create an
additional $2n_++2n_-+2$ opposite-sign charged and $2n_++2n_-$
same-sign charged dipoles.
Due to the separation of scales in our strongly-ordered parton
shower, however, and the fact that the intermediate photon or
gluon is neutral, $2n_++2n_-$ of these dipoles approximately cancel
pairwise, leaving the $2$ newly created dipoles in the
fermion-antifermion system effectively separated from the
dipoles of the system that emitted the photon or gluon
in the first place.
Likewise, since these $f\bar{f}$-pairs are usually produced
near-collinearly, soft photon emissions will not resolve their
individual charges if emitted outside the cone spanned by the
dipole; this is the \QED analogue of angular ordering.
We thus follow the strategy outlined in \cite{Flower:2026byh},
limiting our dipoles to the primary set on one side, and the
newly created dipole-pair on the other side of the split
photons and gluons.
This minimises the number of spectators and the impact of
negative weights while keeping the radiation pattern intact.
A similar stategy was adopted in \Pythia \cite{Bierlich:2022pfr}.

Conversely, where a photon is the splitting parton, we are free
to choose which partons we allow as spectators.
This freedom arises from the fact that photon splittings have
no soft singularity and are instead purely collinear.
The role of the spectator is solely to absorb the arising
transverse recoil.
Unlike for a gluon splitting, which has exactly two spectators
in the large-$N_c$ limit, one or more spectators to a photon splitting
can be considered, and
the recoil can be distributed arbitrarily among them, provided
we satisfy the constraint on $\kappa_{\ijt,\kt}$.
In all cases in this work, we choose to distribute the
transverse recoil democratically between all spectators,
$\kappa_{\ijt,\kt}=-\frac{1}{n_{\text{spec}}}$ where
$n_{\text{spec}}$ is the chosen number of spectators.
Various approaches to the choice of these spectators have been
implemented, see \cite{Schonherr:2017qcj,Dittmaier:1999mb}.

Finally, the coupling factors $c_{ij,k}^a$ are given by
\begin{equation}
  \begin{split}
    c_{ij,k}^\text{\QCD}
    =
      8\pi\,\alphaS
      \left\{
      \begin{array}{ll}
        \mf{T}_\ijt^2 & q_\ijt\to q_ig_j,\,g_\ijt\to g_ig_j \\
        T_R & g_\ijt\to q_i\bar{q}_j
      \end{array}
    \right.
    \qquad\qquad\text{and}\qquad
    c_{ij,k}^\text{\QED}
    =
      8\pi\,\alpha
      \left\{
      \begin{array}{ll}
        Q_{\ijt}^2 & f_\ijt\to f_i\gamma_j \\
        Q_i^2 & \gamma_\ijt\to f_i\bar{f}_j
      \end{array}
      \right.\;.
  \end{split}
\end{equation}
This leaves us to discuss the ordering parameter $t$ and the
infrared cutoff $t_c$.
While we implement the traditional transverse momentum
$t=\kTsq$ of \cite{Schumann:2007mg} for photon and gluon
emissions, we adopt the virtuality $t=q^2$ for photon and
gluon splittings \cite{Flower:2024cpj}.
The infrared cutoff $t_c$ can in principle be set individually
for each splitting.
A convenient and physically-motivated choice is to choose one
cutoff, $t_{\text{\QCD}}$, for splittings in which at least one
of $\ijt,i,j$ is charged under \QCD and another, $t_{\text{\QED}}$,
for splittings in which they are all colour neutral.
With this choice, we facilitate the transition of all
colour charged particles from the regime of perturbative
\QCD implemented in the parton shower to that of
non-perturbative \QCD parametrised in PDFs and hadronisation models.%
\footnote{
  While this includes initial state $q\to q\gamma$ and
  $\gamma\to q\bar{q}$ splittings, it should also include
  initial state $\gamma\to \ell\bar{\ell}$ splittings if
  the photon is extracted from a composite beam particle
  through a non-perturbative PDF.
  All available proton PDFs, with the exception of
  \cite{Bertone:2015lqa,Buonocore:2020nai}, however,
  neglect leptons in their DGLAP evolution.
  For the processes under consideration in this paper,
  such splittings do not play a role and lepton-induced
  contributions are suppressed by multiple orders of
  magnitude \cite{Bertone:2015lqa}.
}
Pure \QED splittings, on the other hand, remain
perturbative as $t\to 0$.
Furthermore, both radiative energy loss through photon radiation
and photon splittings into electron-positron and
muon-antimuon pairs remain
phenomenologically relevant on much smaller scales,
necessitating much smaller cutoff scales for
these splittings.

We further differentiate between initial and final state
emitters, taking into account the validity ranges of
typical proton PDF sets,
\begin{equation}
  \begin{split}
    \begin{array}{l}
      t_{c,\text{\QCD}}^\text{IS}
      \,=\;
      3\,\GeV^2\\
      t_{c,\text{\QCD}}^\text{FS}
      \,=\;
      1\,\GeV^2
    \end{array}
    \qquad\qquad
    \begin{array}{l}
      t_{c,\text{\QED}}^\text{IS}
      \,=\;
      3\,\GeV^2\\
      t_{c,\text{\QED}}^\text{FS}
      \,=\;
      10^{-6}\,\GeV^2
    \end{array}
  \end{split}
\end{equation}
Because the spectator does not determine the value of $t_c$,
it is possible to have cases where an initial state spectator
is involved in a splitting with scale $t<Q_\text{min}^2$,
the minimum scale at which the PDF can be evaluated.
To avoid this, we always probe the initial-state spectator PDF with
scale $\max(t,t^\text{IS}_{c,a_k})$,
$a_k\in\{\text{\QCD},\text{\QED}\}$,
effectively freezing the PDF
at the cutoff appropriate to the spectator species.

\subsection{\texorpdfstring{\MCatNLO}{MC@NLO} matching of \texorpdfstring{\QCD}{\text{QCD}} and \texorpdfstring{\QED}{\text{QED}} emissions}
\label{sec:methods:mcatnlo}

To reach full NLO accuracy, the parton shower constructed
above needs to be matched to the respective exact
$\order(\alpha_s)$ and $\order(\alpha)$ expressions.
To this end, we implement the \MCatNLO method
\cite{Frixione:2002ik}
in the formulation of \cite{Hoeche:2011fd}.
Not only is this formulation of the \MCatNLO method adapted
to matching a parton shower based on Catani-Seymour dipoles
to an NLO calculation performed using a Catani-Seymour subtraction
\cite{Catani:1996vz,Catani:2002hc,Schonherr:2017qcj},
but it also natively accounts for the occurrence of non-positive
definite splitting functions that are ubiquitous in \QED dipoles,
see section \ref{sec:methods:ps} above.

To construct the \MCatNLO master equation, we start from
the fully colour- and spin-dependent
parton shower of Eq.\ \eqref{eq:methods:ps:psfunc}.
We replace
the splitting kernels of Eqs.\ \eqref{eq:methods:ps:pskernels}
and \eqref{eq:methods:ps:pskernels_qcd_qed} explicitly
with the respective Catani-Seymour dipoles,
\begin{equation}\label{eq:methods:mcatnlo:psfunc}
  \bar{\mc{F}}_n(t_n;O)
  =
    \bar{\Delta}_n(t_n,t_c)\,O(\Phi_n)
    +\int_{t_c}^{t_n}
     \done\Phi_1\,
     \frac{\mr{D}_n(\Phi_n\!\!\cdot\!\Phi_1)}{\mr{B}_n(\Phi_n)}\,
     \bar{\Delta}_n(t_n,t_{n+1})\,
     \mc{F}_{n+1}(t_{n+1};O)\;,
\end{equation}
where the barred quantities denote the implicit replacement
$\mr{K}_n \to \mr{D}_n/\mr{B}_n$.
As before, a characteristic scale $t_n$ of the $n$-parton
process acts as the starting scale, limiting the
resummation phase space.
If an emission occurs, the evolution continues with
the standard shower of Eq.\ \eqref{eq:methods:ps:psfunc} on
the newly-formed $(n+1)$-particle state.
The Catani-Seymour dipoles, summing over both \QCD and \QED
splittings,
\begin{equation}\label{eq:methods:mcatnlo:psdipoles_qcd_qed}
  \mr{D}_n(\Phi_n\!\!\cdot\!\Phi_1)
  =
    \sum_{a\in\{\text{\QCD,\QED}\}}
    \mr{D}_n^a(\Phi_n\!\!\cdot\!\Phi_1)
  =
    \sum_{a\in\{\text{\QCD,\QED}\}}
    \sum_{\{i,j,k\}}
    \mr{D}_{ij,k}^a(\Phi_n\!\!\cdot\!\Phi_1^{ij,k})
  \,,
\end{equation}
are defined on the factorised $(n+1)$-parton phase space,
$\Phi_{n+1}=\Phi_n\!\!\cdot\!\Phi_1$ with $\Phi_1\equiv\Phi_1(t,z,\phi)$
as above, with the emitter, emittee, and spectator indices
$i$, $j$, and $k$ as before.
The additional superscripts in $\Phi_1^{ij,k}$ indicate explicitly
that the phase space factorisation is dipole-dependent,
whereas this dependence was left implicit in section \ref{sec:methods:ps}
above.\footnote{
  For simplicity, we have elected to use the notation of final-final
  dipoles throughout, however this should be understood to implicitly
  include all four dipole types shown in Fig.\ \ref{fig:methods:ps:dips}.
}
The corresponding Sudakov form factor is defined likewise,
\begin{equation}\label{eq:methods:mcatnlo:pssudakov}
  \bar{\Delta}_n(t,t')
  =
    \exp\left[
      -\int_{t'}^{t}\done\Phi_1\,
       \frac{\mr{D}_n(\Phi_n\!\!\cdot\!\Phi_1)}{\mr{B}_n(\Phi_n)}
    \right]
  =
    \prod_{a\in\{\text{\QCD,\QED}\}}\!\!\!\!
    \bar{\Delta}_n^a(t,t')
    \,.
\end{equation}
As a result, like its conventional parton shower analogue in
the previous section, the fully colour- and spin-dependent
parton shower generation functional of Eq.\ \eqref{eq:methods:mcatnlo:psfunc}
implements an interleaved \QCD and \QED evolution that is not
the simple sum of either evolution.

With the help of the dipole terms and Sudakov factor,
we can now express the NLO-accurate expectation value of an
arbitrary infrared-safe observable as
\begin{equation}\label{eq:methods:mcatnlo:mcatnlo_obs}
  \langle O \rangle^{\text{\MCatNLO}}
  =
    \int\done\Phi_n\;
    \bar{\mr{B}}(\Phi_n)\,\bar{\mc{F}}_n(t_n;O)
    +\int\done\Phi_{n+1}\;
    \mr{H}_n(\Phi_{n+1})\,\mc{F}_{n+1}(t_{n+1};O)
    \,.
\end{equation}
Therein, the $\bar{\mr{B}}$-function
\begin{equation}\label{eq:methods:mcatnlo:Bbar}
  \begin{split}
    \bar{\mr{B}}_n(\Phi_n)
    \,=\;&
      \mr{B}_n(\Phi_n)
      +\sum_{a\in\{\text{\QCD,\EW/\QED}\}}
      \left[
        \tilde{\mr{V}}_n^a(\Phi_n)
        +\int_0^{t_n}\done\Phi_1\,\mr{D}_n^a(\Phi_n\!\!\cdot\!\Phi_1)
        \right]
  \end{split}
\end{equation}
collects the Born contribution, $\mr{B}_n$, along with
the renormalised \QCD and \EW virtual corrections,
$\tilde{\mr{V}}_n^\text{\QCD}$ and $\tilde{\mr{V}}_n^\text{\EW}$,
which include the collinear counterterms of the PDF factorisation,
as well as the integral of the dipole splitting functions
over the parton shower's single emission phase space.
Events generated by this term are referred to 
as $\mr{S}$-events.

Furthermore, the $\mr{H}$-events correct the first emission
to the exact NLO real-correction emission pattern
\begin{equation}\label{eq:methods:mcatnlo:H}
  \begin{split}
    \mr{H}_n(\Phi_{n+1})
    \,=\;&
      \sum_{a\in\{\text{\QCD,\QED}\}}
      \mr{H}_n^a(\Phi_{n+1})
    =
      \sum_{a\in\{\text{\QCD,\QED}\}}
      \left[\mhl
        \mr{R}_{n+1}^a(\Phi_{n+1})
        -\mr{D}_n^a(\Phi_n\!\!\cdot\!\Phi_1)\Theta(t_n-t_{n+1})
      \right]\,,
  \end{split}
\end{equation}
using the \QCD and \QED real emission corrections, $\mr{R}_n^\text{\QCD}$
and $\mr{R}_n^\text{QED}$.
With these definitions, expanding
Eq.\ \eqref{eq:methods:mcatnlo:mcatnlo_obs} to $\order(\alphaS)$ and
$\order(\alpha)$, respectively, demonstrates its NLO \QCDpEW accuracy.
It is important to note that the $(n+1)$-parton state, created
either through the modified parton shower of
Eq.\ \eqref{eq:methods:mcatnlo:psfunc} for an $\mr{S}$-event or
directly through an $\mr{H}$-event, continues its evolution
by means of the ordinary parton shower $\mc{F}_{n+1}$ of the
previous section, using $t_{n+1}$ as its starting scale.\footnote{%
  More precisely, the starting scale is given by 
  $\min{(t_n,t_{n+1})}$, since it is possible for $\mr{H}$-events
  to have $t_{n+1}<t_n$ and we wish to avoid to avoid extending 
  the resummation beyond its region of validity.
  For further discussion, see \cite{Flower:2026byh}.
}

\paragraph*{Enhancing \texorpdfstring{\QED}{\text{QED}} splittings.}
In an interleaved \QCDpQED shower evolution we expect that
the \QCD splittings we generate greatly outnumber the \QED 
splittings, both due to their larger coupling and presence of
the $g\to gg$ channel that is doubly enhanced (in addition to
the larger coupling constant, $C_A$ is large compared to the
average $Q_{\ijt,\kt}^2$).
Evolving the interleaved shower, we expect to see
approximately the same number of each type of splitting (\QCD or \QED)
as we would see if the shower evolved only in \QCD or only in \QED,
respectively.\footnote{
  To be precise, we should see slightly increased numbers of both types
  of splitting in the interleaved
  shower because quarks can be produced in either photon or gluon 
  splittings and, in turn, emit either type of radiation.
}
However, if we truncate the shower after a fixed number of emissions,
\QCD splittings will almost always outcompete \QED ones.
This can lead to much worse statistics in photon-based observables
but is of particular concern in the context of matching.
NLO matching in our formulation is designed to correct the
first shower emission,
which is overwhelmingly likely to be a \QCD splitting,
to its corresponding real matrix element.
It follows that, if we rarely generate a \QED splitting as the first
emission, we may not correctly match to NLO \EW using finite statistics.

This is especially problematic for multiplicative matching algorithms
that rely on the parton shower to fill the real emission phase space.
For \MCatNLO, on the other hand,
the $\mr{H}$-events always generate both \QCD and \QED
real emissions according to their respective matrix elements,
regardless of the first shower emission.
For the subtraction to be valid, however, we must ensure that enough
\QED splittings occur in the \MCatNLO shower, and that we use identical
dipole splitting functions for the entire process.
In particular, if the later shower uses different dipoles from the
first emission,
\QED splittings that did not occur first would not correctly
cancel the dipoles subtracted from \QED $\mr{H}$-events.

It can therefore be useful, even in an \MCatNLO calculation, to
artificially enhance the number of \QED splittings we generate
in the shower and correct for this in the event weight.
This ensures the reliability of the matching and can help to improve
the statistical convergence of photon-based observables,
which otherwise have poor sample sizes,
although a balance must be struck between increasing sample size
and minimising variation of event weights.
To this end, we make use of the weighted veto algorithm
as outlined in \eg App.\ B of \cite{Hoeche:2009xc}.

\paragraph*{Treatment of H-events.}
The different magnitudes of \QCD and \EW corrections
leads to noteworthy consequences in the $\mr{H}$-events of
our \MCatNLO \QCDpEW calculation.
This can be illustrated exemplarily by considering an observable,
$O^\gamma$ that studies the hardest photon emission,
\eg its transverse momentum.
Following Eq.\ \eqref{eq:methods:mcatnlo:mcatnlo_obs}, and
keeping only terms for which $O^\gamma\neq 0$, we find
\begin{equation}\label{eq:methods:mcatnlo:Ogamma}
  \begin{split}
    \langle O^\gamma\rangle
    \,=\;&
      \int\done\Phi_n\done\Phi_1\;
      \left\{
        \mr{R}_n^\text{\QED}
        +
        \left[
          \frac{\bar{\mr{B}_n}(\Phi_n)}{\mr{B}_n(\Phi_n)}\,
          \bar{\Delta}_n(t_n,t_{n+1})
          -1
        \right]
        \mr{D}_n^\text{\QED}(\Phi_n,\Phi_1)
      \right\}
      O^\gamma(\Phi_{n+1})
      +\ldots
      \;.
  \end{split}
\end{equation}
While this expression describes the leading contribution,
starting at $\order(\alpha)$, where the leading photon is
generated by our interleaved \MCatNLO directly,
the ellipsis contains additional contributions,
starting at $\order(\alpha_s\alpha)$
where the \MCatNLO produces a \QCD emission first
and the leading photon is emitted in the ensuing
standard parton shower evolution.
While the inner square bracket in
Eq.\ \eqref{eq:methods:mcatnlo:Ogamma}, expanded to fixed-order,
vanishes to the LO accuracy our \MCatNLO \QCDpEW is
formally expected to have, giving the desired result,
large logarithms can spoil this finding.
To be precise, the Sudakov factor this bracket contains,
driven by its \QCD dynamics, is numerically small
compared to unity, $\bar{\Delta}\ll 1$, leading to
a numerically large higher-order effect that can
lead to artefacts in the physical spectrum,
including $O^\gamma<0$ in some regions of phase space.

We mitigate this problem by applying a similar Sudakov
factor, computed through the standard interleaved
parton shower of Sec.\ \ref{sec:methods:ps},
to the $\mr{H}$ events in order
to reduce the magnitude of the negative contribtution
to Eq.\ \eqref{eq:methods:mcatnlo:Ogamma}.\footnote{
  Albeit in different contexts, a similar strategy was proposed
  in the original formulation of merging multiple \MCatNLO calculations
  of successive multiplicities (\MEPSatNLO)
  \cite{Hoeche:2012yf,Gehrmann:2012yg},
  and in the context of negative weight reduction from $\mr{H}$-events
  in \cite{Frederix:2020trv,Danziger:2021xvr}.
}
This changes the definition of the $\mr{H}$ events of
Eq.\ \eqref{eq:methods:mcatnlo:H} to
\begin{equation}\label{eq:methods:mcatnlo:Hsud}
  \begin{split}
    \mr{H}_n^\text{sud}(\Phi_{n+1})
    \,=\;&
      \mr{H}_n(\Phi_{n+1})\;
      \Delta_n(t_n,t_{n+1})
      \,.
  \end{split}
\end{equation}
While this choice introduces terms beyond
our formal accuracy, it serves to reduce the unphysical
effects of the already present higher-order contributions.
With this change, Eq.\ \eqref{eq:methods:mcatnlo:Ogamma}
becomes
\begin{equation}\label{eq:methods:mcatnlo:Ogamma_sud}
  \begin{split}
    \langle O^\gamma\rangle
    \,=\;&
      \int\done\Phi_n\done\Phi_1
      \left\{
        \mr{R}_n^\text{\QED}(\Phi_n\!\!\cdot\!\Phi_1)
        +
        \left[
          \frac{\bar{\mr{B}}_n(\Phi_n)}{\mr{B}_n(\Phi_n)}\,
          \frac{\bar{\Delta}_n(t_n,t_{n+1})}{\Delta_n(t_n,t_{n+1})}
          -1
        \right]
        \mr{D}_n^\text{\QED}(\Phi_n,\Phi_1)
      \right\}
      \Delta_n(t_n,t_{n+1})\,
      O^\gamma(\Phi_{n+1})
      \\
    &{}+\ldots
      \,.
      \\[-2mm]
  \end{split}
\end{equation}
As is evident, in regions where the Sudakov form factor
becomes small by logarithmically large \QCD corrections,
these numerically large corrections cancel well
as long as both Sudakov factors are similar enough, \ie as
long as the leading-colour, spin-averaged parton shower
approximation works well.

However, it is important to note that the modification
of Eq.\ \eqref{eq:methods:mcatnlo:Hsud} changes the
inclusive cross section by a term of $\order(\alpha_s\alpha)$.
Whilst their inclusive impact is small, events dominated
by $t_{n+1}\ll t_n$ $\mr{H}$-event topologies akin to
$O^\gamma$ are affected signifantly.
The introduced Sudakov factors, however, are the dominant
$\order(\alpha_s\alpha)$ corrections expected in that region
\cite{Hoeche:2012yf,Gehrmann:2012yg}.
Further, as a by-product, the impact of the negative weights
that often arise from $\mr{H}$-events is reduced
\cite{Frederix:2020trv,Danziger:2021xvr}.
We refer to calculations performed using this additional
Sudakov factor as \MCatNLOsud.

\subsection{Resonance aware subtraction and parton shower evolution}
\label{sec:methods:resaware}

In this section, we discuss embedding
processes with internal resonances into our \QCDpQED
matched calculation and subsequent parton evolution.

\begin{figure}[t!]
  \includegraphics[width=0.45\textwidth]{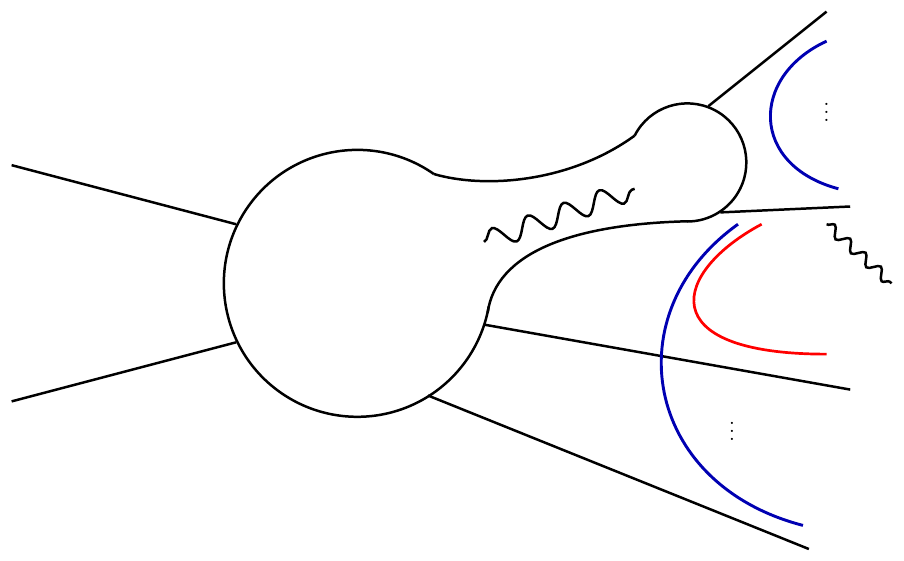}
  \hfill
  \includegraphics[width=0.45\textwidth]{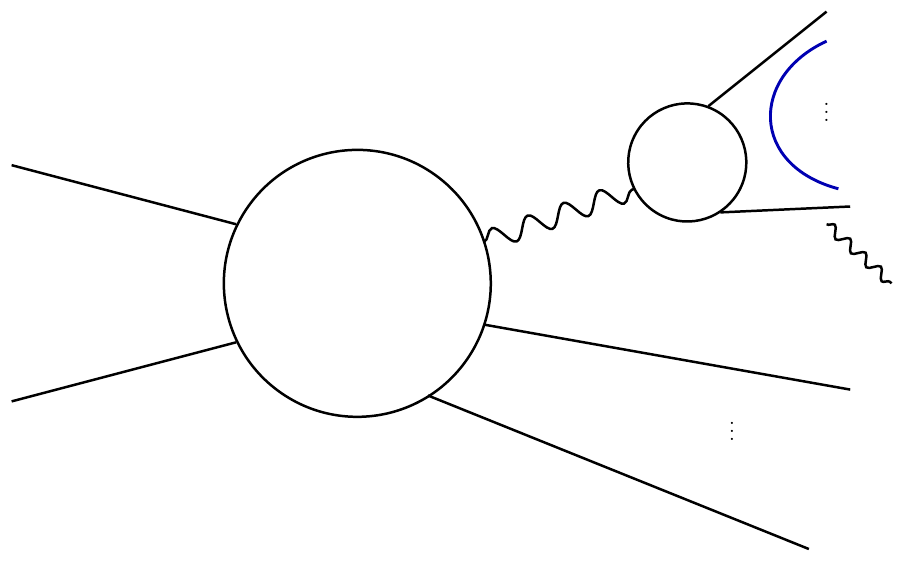}
  \caption{%
    Diagrams of standard (left) and resonant-aware (right)
    subtraction terms and parton shower kernels.
    \label{fig:methods:ressub:diag}
  }
\end{figure}

\paragraph*{Resonance aware subtraction.}
Dipole factorisation in the soft and collinear limits
assumes that the leading change in the matrix element
originates in the near on-shell propagator of the emitting
parton, whereas the underlying lower-order matrix element
remains approximately constant \cite{Catani:1996vz,Catani:2002hc}.
In the presence of resonances, $s$-channel propagators
of unstable massive particles with a finite width,
this assumption no longer holds when the characteristic
scale of the emission\footnote{%
  More precisely, the change in the resonance's
  virtuality induced by the momentum transfer to
  the spectator, which is intimately linked
  with the characteristic scale of the emission.
} is of the order of the width of the resonance.
Now, when the virtuality of the resonant propagator
in the lower-order process is near its nominal mass,
the dominant change in the (sufficiently hard) emission
amplitude originates in the shift of the propagator
virtuality away from that nominal mass 
\cite{Khoze:1992rq,Dokshitzer:1992nh,Khoze:1998ap}.
In this situation, the standard dipole subtraction
is inappropriate and a formulation that incorporates
resonance information must be used.

The effect of a resonance in the underlying
matrix element can be incorporated in the subtraction formalism
by considering the ability of an emission to resolve it,
see Fig.\ \ref{fig:methods:ressub:diag},
akin to the strategy followed in 
\cite{Jezo:2015aia,Frederix:2016rdc,Denner:2026ztd,Hoche:2026dup}.
For emissions in the soft and/or collinear limit, where their
emission scale is small compared to the width of the
resonance (and thus their wavelength too long to resolve it),
only the external asymptotic states are resolved.
All dipoles that would have been present without the resonance
are active in the subtraction.
In the dipole picture, the external states act as emitters and spectators
and the recoil transferred across the resonant propagator
only changes its virtuality minutely, fulfilling the basic
factorisation assumptions.
Away from these limits, when the emission scale is
much larger than the width of the resonance,
the emission can resolve the resonance
and the process factorises into resonance production
and resonance decay, which act as separate subprocesses.
In this case, dipoles are localised in either production or
decay, and dipoles connecting the two are suppressed by
a factor of $\tfrac{\Gamma}{m}$ \cite{Khoze:1998ap}.

To this end, we introduce two criteria to characterise
the phase space in which an emission is ``aware'' of
the presence of a resonance, and thus the resonance-aware
subtraction is used instead of the standard subtraction.
\begin{itemize}
  \item[$\boldsymbol{\Deltares}$]
    The factorisation of the process into
    resonance production and decay subprocesses for hard
    emissions only holds if the virtuality of the resonant
    propagator is near its nominal mass shell.
    Hence, we define the measure $\Delta_r = \left|\sqrt{s_r}-m_r\right|\!/\Gamma_r$ \cite{Kallweit:2017khh},
    where $\sqrt{s_r}$ is the virtuality of the resonant propagator,
    and $m_r$ and $\Gamma_r$ are its nominal mass and width.
    An internal propagator can then be defined as
    resonant if $\Delta_r<\Deltares$.
    Although the precise distinction between resonant
    and non-resonant topologies is arbitrary, values
    of $\Deltares=2\ldots 10$ have been
    found reasonable \cite{Sherpa:2024mfk}.
  \item[$\boldsymbol{\tres}$]
    Only emissions with $t_{n+1}\gg \Gamma_r^2$
    are hard enough to resolve the internal resonance, while
    emissions with $t_{n+1}\ll \Gamma_r^2$ are too soft
    to resolve it.
    We thus choose a scale $\tres$ of the order of the
    resonance width $\Gamma_r^2$ to delineate the
    production-decay-factorised ($t_{n+1}>\tres$) 
    and -non-factorised regimes ($t_{n+1}<\tres$).
\end{itemize}
It is essential to apply both criteria equally to the
forward-evolved $\mr{S}$-events, and the $\mr{H}$-events
where the $(n+1)$th parton is already present, such that
the subtraction is exact and does not introduce spurious
divergences.
Thus, for all subtraction dipoles, most notably those in
the $\mr{H}$-events, $\Delta_r$ is always calculated from
the underlying Born configuration.

Fig.\ \ref{fig:methods:ressub:diag} sketches the dipole
configurations in both the production-decay-factorised and
non-factorised regimes.
While the production-decay-factorised
regime is, by construction, infrared finite, the
inclusion of its separate subtraction is still beneficial
in view of numerical stability and necessary when
matched to a parton shower evolution.

The above definitions
can be straightforwardly applied to processes with multiple and
competing resonances by ordering possible resonant structures
in ascending $\Delta_r$ akin to \cite{Kallweit:2017khh}.

\paragraph*{Collinear Casimir factors.}
When using Catani-Seymour dipoles, care has to be taken
to preserve the collinear limit of the respective splitting
processes.
Their collinear Casimir factor ($\mf{T}_\ijt^2=\mf{T}_i^2$
for $\ijt\to ig$ and $T_R$ for $g\to q\bar{q}$ in \QCD splittings,
and $Q_\ijt^2=Q_i^2$ for $\ijt\to i\gamma$ and $Q_i^2$ for
$\gamma\to ij$ in \QED splittings) is distributed
over all dipoles with all possible spectators $\kt$,
see Eqs.\ \eqref{eq:methods:ps:Cijk} and \eqref{eq:methods:ps:Qijk}.
Factorisation into resonance production and decay
along a (colour) neutral resonance implies that the sum
of the $\mf{C}_{ij,k}^\text{\QCD/\QED}$ of all dipoles
that span across the resonance vanishes due to (colour)
charge conservation.
Thus, the sum of (colour) charge
factors of the subset of dipoles local to either the
production or decay subprocess produces the correct
collinear Casimir factor.
(Colour) charged resonances, however, like
the top quark in \QCD and the $W$ boson in \QED, necessitate
the addition of resonant dipoles akin to \cite{Basso:2015gca}
in the factorised regime.

\begin{figure}[t!]
  \centering
  \includegraphics[width=0.35\textwidth]{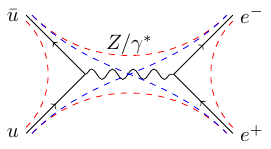}
  \hspace*{0.1\textwidth}
  \includegraphics[width=0.35\textwidth]{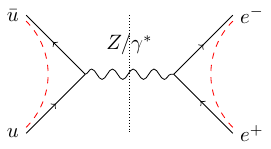}
  \caption{
    Diagrams of standard (left) and resonant-aware (right)
    subtraction terms and parton shower kernels in Drell-Yan
    production in the quark-initiated channels
    with positive dipoles shown in red and negative in blue
    (the initial-final/final-initial dipoles would have the 
    opposite sign for $d$-type quarks).
    The regimes are separated by the emission scale $\tres$
    and $Z$ boson virtuality $\Deltares$.
    The photon initiated channel is non-resonant
    and, thus, is not factorised at $(\Deltares,\tres)$.
    \label{fig:methods:resaware:diag_dy}
  }
\end{figure}

\paragraph*{Explicit construction.}
Focusing on the process under consideration in this
paper, inclusive production of a lepton pair at the LHC
at \MCatNLO \QCDpEW, we set up our subtraction as follows.
While the subtraction of all \QCD divergences proceeds
as in the original Catani-Seymour dipole formulation---all
\QCD dynamics are localised in the resonance
production subprocess irrespective of its virtuality---the
subtraction of \QED infrared divergences includes information
on the internal resonance, as illustrated in
Fig.\ \ref{fig:methods:resaware:diag_dy}.

In addition to the standard non-resonant-aware \QCD and \QED
Catani-Seymour dipole subtraction terms $\mr{D}_{ij,k}^a$,
we introduce the following resonance-aware
final-initial and initial-final real-subtraction dipoles
that span the (colour) neutral $Z$ boson resonance,
\ie where emitter $\ijt$ and spectator $\kt$ are at different
ends of the resonant propagator,
\begin{equation}\label{eq:methods:resaware:resdips}
  \begin{split}
    \mr{D}_{ij,k}^{a,\text{res}}
    \,=\;&
      \mr{D}_{ij,k}^a\;
      \left[\,
        \Theta(\tres-t)\,
        \Theta(\Deltares-\Delta_r)
        +
        \Theta(\Delta_r-\Deltares)\,
      \right]
    \;,
  \end{split}
\end{equation}
where $t$ is the emission phase space parameter of
Eq.\ \eqref{eq:methods:ps:psfunc}.
The set of $\Theta$-functions switch these resonance-spanning
dipoles off in the regime where the emission can resolve
the resonance according to $\Deltares$ and $\tres$.
All other dipoles are kept in their non-resonance-aware
form.
As discussed, new dipoles that include the resonance as
emitter or spectator are not necessary for
neutral-current Drell-Yan.

This form of $\mr{D}_{ij,k}^\text{res}$ necessarily
introduces logarithms of the form
$\log\frac{\tres}{t_n}\sim \log\frac{\tres}{m_r^2}$
in the subtraction \cite{Jezo:2015aia}; however, these only
arise in the width-suppressed interference terms so they appear as
$\frac{\Gamma_r}{m_r}\log\frac{\tres}{m_r^2}$, which is finite as
$\Gamma_r\to 0$ for $\tres \sim \Gamma_r^2$.
The difference of the $\mr{D}_{ij,k}^\text{res}$ and
the original $\mr{D}_{ij,k}$ is therefore finite in four dimensions,
so we compute their integrated counter-parts needed for
Eq.\ \eqref{eq:methods:mcatnlo:Bbar} numerically,
\begin{equation}\label{eq:methods:resaware:intdips}
  \begin{split}
    \int_0^{t_n}\done\Phi_1\,\mr{D}_n^{a,\text{res}}(\Phi_n\!\!\cdot\!\Phi_1)
    \,=\;&
      \int_0^{t_n}\done\Phi_1\,\mr{D}_n^a(\Phi_n\!\!\cdot\!\Phi_1)
      +\int_0^{t_n}\done\Phi_1
       \left[\mhl
         \mr{D}_n^{a,\text{res}}(\Phi_n\!\!\cdot\!\Phi_1)
         -\mr{D}_n^{a}(\Phi_n\!\!\cdot\!\Phi_1)
       \right]
    \;,
  \end{split}
\end{equation}
using the techniques of \cite{Hoeche:2012fm}.

With these definitions at hand, we can now straightforwardly
compute an $\mr{S}$-event.
First, we generate a Born phase space point and determine
whether this configuration is resonant or non-resonant
according to $\Deltares$.
Its $\bar{\mr{B}}$-function,
including the numerically integrated resonance-aware dipoles
of Eq.\ \eqref{eq:methods:resaware:intdips}, is calculated
accordingly.
The matched full-colour spin-dependent emission is generated
according to $\bar{\mc{F}}_n$ of
Eq.\ \eqref{eq:methods:mcatnlo:psfunc}, again using both
standard and resonance-aware dipoles.
From here on, the resonance-aware dipole shower discussed
below continues the evolution.

For $\mr{H}$-events, on the other hand, determining $\Delta_r$
and $t_{n+1}$ needs more care.
We begin by generating a real phase space point;
for each dipole, we map this to an underlying Born configuration
which allows us to compute $\Delta_r$ and the splitting scale $t_{n+1}$
and thus determine whether the dipole is active.
Each surviving dipole is a possible one-step parton shower
history for the real configuration.
Continuing the parton shower evolution requires a single scale
$t_{n+1}$ and resonance measure $\Delta_r$ to be assigned to
the entire $\mr{H}$-event, so we must cluster the real emission
according to the most likely shower history
\cite{Hoche:2026dup,Hoeche:2012yf}.
The weight of each potential clustering should be the
shower-generated matrix element, \ie the underlying Born
multiplied by the emission probability, however it is
computationally expensive to use the exact matrix element so we
instead approximate it.
For resonant processes, we take the Born to be proportional to the resonant
Breit-Wigner function $\left[(q_r^2-m_r^2)^2+m_r^2\Gamma_r^2\right]^{-1}$
whereas for non-resonant processes, we return to our assumption
that the dominant change in the matrix element arises from
the emitting propagator, captured by the splitting kernel,
and approximate the Born as constant, with continuity
at $\Delta_r=\Deltares$.
For the emission probability, we use the large-$N_c$ spin-averaged
splitting functions of Eq.\ \eqref{eq:methods:ps:pskernels_qcd_qed}.
The weight for each clustering is then given by
\begin{equation}
  \mr{S}_{ij,k}^a
  \,=\;
  \begin{cases}
    \qquad\;\;
    \frac{\mr{K}_{ij,k}}{(q_r^2-m_r^2)^2+m_r^2\Gamma_r^2}
      \qquad\qquad\;\,\,\Delta_r<\Deltares \\[1ex]
    \frac{\mr{K}_{ij,k}}{(\Deltares\Gamma_r(2m_r+\Deltares\Gamma_r))^2+m_r^2\Gamma_r^2}
      \qquad\Delta_r>\Deltares
      \,,\;
      q_r^2>m_r^2 \\[1ex]
    \frac{\mr{K}_{ij,k}}{(\Deltares\Gamma_r(2m_r-\Deltares\Gamma_r))^2+m_r^2\Gamma_r^2}
      \qquad\Delta_r>\Deltares
      \,,\;
      q_r^2<m_r^2
  \end{cases}
  \;
\end{equation}
and we choose the largest weight in a `winner takes all' approach.
Note that the Breit-Wigner factor is applied regardless of
the value of $t_{n+1}$,
since the ability of the emission to resolve the resonance does not change
the approximate shape of the matrix element itself.
Having used this clustering to identify the scale of the real
emission, $t_{n+1}$ we
calculate the $\mr{H}$-event weight.
Like for the $\mr{S}$-events, the resonance-aware dipole
shower described below continues the evolution from this scale.

\paragraph*{Dipole showers in the presence of resonances.}

Beyond \MCatNLO, the parton shower requires resonance-aware
dipoles for the assembly of splitting kernels to reconstruct
the correct infrared structure of the $(n+k+1)$-th emission
off the $(n+k)$-parton configuration.
Hence, we keep the dipole configurations described above
within the resonant regime delineated by $\Deltares$ and $\tres$.
Once the parton shower evolution has left the resonant
regime, the full non-resonant structure is used. The
transition is similarly implemented using $\Theta$-functions
as defined in Eq.\ \eqref{eq:methods:resaware:resdips}.
For the process under consideration, in addition to the
initial-final and final-initial dipoles of the first emission which
connect emitter-spectator pairs in the production and decay
of the $Z$-boson resonance, we now also have to consider
resonance-aware final-final dipoles connecting the emissions
from the production subprocess, now in the final state, with
partons in the decay subprocess.
We define them in the same manner following
Eq.\ \eqref{eq:methods:resaware:resdips}, \ie these dipoles
are only active
for emissions at scales $t<\tres$ or $\Delta_r>\Deltares$.

A new feature that only enters in the parton shower evolution
beyond the matched first emission for our resonant process is
photon splittings.
Here, we enforce spectators to reside on the same side of the
resonance at all scales.
While the production of new \QCD particles is possible in the
decay subprocess of the resonance through $\gamma\to q\bar{q}$
splittings of photon emissions off the final state leptons,
the parton shower's large-$N_c$ limit enforces that they are
never colour-connected to any parton in the production subprocess,
and hence, resonance-spanning dipoles are not encountered.

While the above is sufficient to preserve the resonance,
the resonance-aware \QED dipole shower suffers the same
degrading statistical performance in the later stages
of the shower through the proliferation of particle-antiparticle
pairs as the standard \QED dipole shower described in
Sec.\ \ref{sec:methods:ps}.
We choose to address this using the same single-spectator scheme,
and restricting dipoles in this way simultaneously prevents charged
emitters from forming dipoles across our (colour-)neutral resonance.
In this case, the transition thresholds have no effect on the shower
as the single-\QED-spectator scheme is used for the entire shower
evolution.

\section{Results}
\label{sec:results}

In this section, we start by demonstrating the validity
of the methods detailed in the previous section,
before presenting our precision calculation for lepton-pair
production in the vicinity of the $Z$-boson resonance at the
LHC in Sec.\ \ref{sec:results:prediction}.

\paragraph*{Input parameters.}
Throughout, we use the PDF set \texttt{NNPDF31\_nlo\_as\_0118\_luxqed}
\cite{Bertone:2017bme}
interfaced through \LHAPDF \cite{Buckley:2014ana}.
This PDF set works in the $n_f=5$ massless quark flavour
scheme and contains a parametrisation of the strong coupling
with $\alphaS(m_Z)=0.118$ in the \MSbar scheme.
Importantly, this includes the dominant QED contribution
to parton evolution, excluding $\gamma\to\ell\bar{\ell}$
splittings, and thus introduces a photon distribution alongside
the quark and gluon distributions.
In the initial-state parton evolution, we must respect the
validity region of this PDF set, $Q^2>2.7 \,\GeV^2$.
In addition, we include all contributing partonic channels
supported by this PDF set throughout, including photon-initiated
channels where applicable.

We work in the complex mass scheme \cite{Denner:2005fg},
\begin{equation}
  \label{eq:results:cms}
  \begin{split}
    \mu_i^2
    \,=\;&
    m_i^2 - im_i\Gamma_i
    \qquad\qquad\text{with}\qquad\qquad
    i\in\{W,Z,H,t\}\;,
  \end{split}
\end{equation}
with the following masses and widths
\begin{equation}
  \begin{array}{rcllcrcll}
    m_W      & \!\!=\!\! & \phantom{0}80.379 & \GeV & \qquad &
    \Gamma_W & \!\!=\!\! & 2.085 & \GeV \\
    m_Z      & \!\!=\!\! & \phantom{0}91.1876\!\!\!\! & \GeV & \qquad &
    \Gamma_Z & \!\!=\!\! & 2.4952\!\!\!\! & \GeV \\
    m_H      & \!\!=\!\! & 125.09 & \GeV & \qquad &
    \Gamma_H & \!\!=\!\! & 0.0041\!\!\!\! & \GeV \\
    m_t      & \!\!=\!\! & 172.5 & \GeV & \qquad &
    \Gamma_t & \!\!=\!\! & 1.32 & \GeV\,,
  \end{array}
\end{equation}
and treat all other flavours as massless.\footnote{%
  To be precise, whilst we work in the $n_f=5$ flavour
  scheme in the
  perturbative hard scattering matrix element
  and the matched first \MCatNLO emission
  (consistent with the chosen PDF set), in the parton
  shower we introduce
  masses for $c$ and $b$ quarks as well as all leptons
  to improve the description of gluon and photon
  radiation and splittings at low scales.
}
We use the $\Gmu$ scheme to renormalise the EW sector
with $(\GF,\mu_W^2,\mu_Z^2)$ as input parameters.
Other appropriate EW input parameter and renormalisation
schemes for the processes under consideration in this
paper are available, including the $\alpha(m_Z)$ and
$\sinweff$ \cite{Chiesa:2019nqb} schemes.
The Fermi constant itself takes the value
\begin{equation}
  \label{eq:results:GF}
  \GF=1.16639\times10^{-5}\,\GeV^{-2}\;.
\end{equation}
In order to match the NLO EW correction to the parton shower,
however, we use the $\alpha(0)$ scheme, defined through the
input parameters $(\alpha(0),\mu_W^2,\mu_Z^2)$, as the coupling
in the photon radiation splitting functions and for the
additional power of the electroweak coupling \cite{Flower:2026byh}.
The two different $\alpha$ values are given by
\begin{equation}
  \label{eq:results:defalpha}
  \begin{split}
    \alpha_{\Gmu}
    \,=\;&
      \left|\frac{\sqrt{2}\,s_\text{w}^2\mu_W^2\GF}{\pi}\right|
    \qquad\qquad\text{and}\qquad\qquad
    \alpha(0) = 1/137.03599976\;,
  \end{split}
\end{equation}
with the weak mixing angle
\begin{equation}
  \label{eq:results:defsw}
  \begin{split}
    s_\text{w}^2
    \,=\;&
      1-c_\text{w}^2
    =
      1-\frac{\mu_W^2}{\mu_Z^2}\;.
  \end{split}
\end{equation}

\paragraph*{Considerations on photon splittings.}
Photon splittings, as discussed in Sec.\ \ref{sec:methods:ps},
allow for some freedom in the choice of spectator.
For the standard subtraction, resonance-aware subtraction, and
the one-emission modified parton shower used in the \MCatNLO
matching of Sec.\ \ref{sec:methods:mcatnlo}, we allow initial
state photon splittings to have the
other initial-state particle as a spectator.\footnote{
  We make this choice even if the other initial-state
  particle is a gluon and does not interact under \QED,
  although this case does not arise in the processes under
  consideration here.
}
In principle, the same argument applies to final-state photon
splittings, but the consideration of processes with Born-level
final-state photons is beyond the scope of this paper.
In the remaining parton shower evolution,
when no resonance is identified we allow all charged
partons to be spectators to photon splittings.
In the resonance-aware evolution, we restrict the allowed set
of spectators to originate from the same side of the resonance
as the splitting photon.

\paragraph*{Implementation.}
We implement all calculations discussed in this paper in the
Monte-Carlo event generator \Sherpa \cite{Sherpa:2024mfk}.
We employ the matrix element generators \Amegic \cite{Krauss:2001iv}
and \Comix \cite{Gleisberg:2008fv} for all tree-level contributions,
including the automated subtraction of all \QCD and \QED infrared
divergences \cite{Schonherr:2017qcj,Gleisberg:2007md}.
All \QCD and \EW renormalised one-loop corrections are
provided by \OpenLoops
\cite{Cascioli:2011va,Kallweit:2014xda,Buccioni:2019sur}
using \Collier \cite{Denner:2016kdg},
\CutTools \cite{Ossola:2007ax},
and \OneLoop \cite{vanHameren:2010cp}.
Events are then parton showered using the
\CSShower \cite{Flower:2026byh,Schumann:2007mg,Hoeche:2009xc}.
Though easily incorporated, we omit any further non-perturbative
\QCD  effects, \ie hadronisation, hadron decays, and multi-parton
interactions, in order to focus on the perturbative \QCD and \EW
properties of the calculations under consideration in this paper.

All analyses are performed using Rivet \cite{Buckley:2010ar,Bierlich:2024vqo}.

\subsection{Validation}
\label{sec:results:validation}

\begin{figure}[t!]
    \centering
    \includegraphics[width=0.49\textwidth]{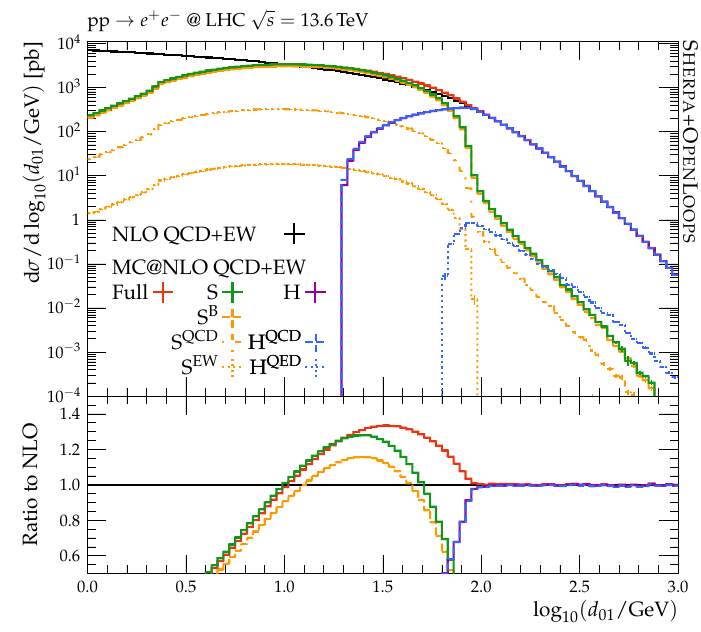}
    \includegraphics[width=0.49\textwidth]{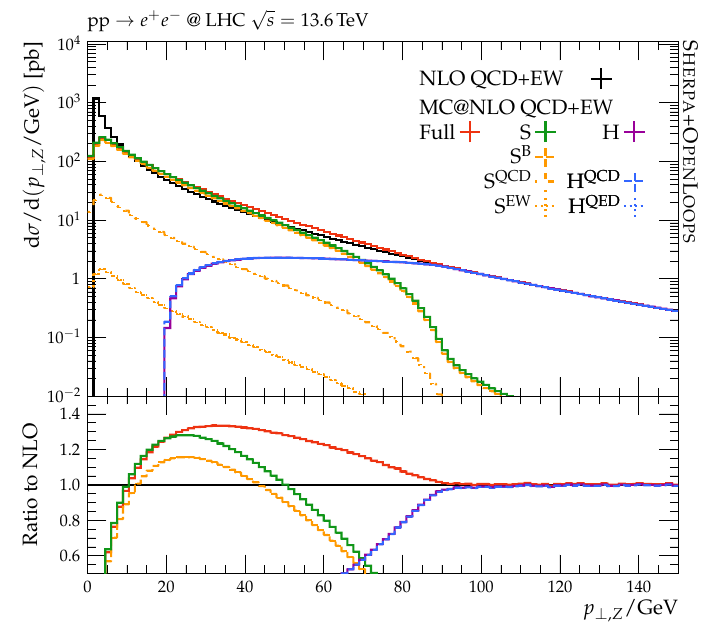}
    \caption{
      Observables for $\mr{pp}\to \nu\bar{\nu}$ at NLO \QCDpEW
      compared to single-emission \MCatNLO \QCDpEW with $\mr{S}$- and $\mr{H}$-events shown separately.
      \MCatNLO events are further separated according to the term of
      Eqns.\ \eqref{eq:methods:mcatnlo:Bbar} and \eqref{eq:methods:mcatnlo:H}
      that they are generated from.
      For convenience, we have defined
      $\mr{S}^\mr{B}=\mr{B}_n\otimes\bar{\mc{F}}_n$
      and $\mr{S}^a=
            \left[
                \tilde{\mr{V}}_n^a(\Phi_n)
                +\int_0^{t_n}\done\Phi_1\,\mr{D}_n^a(\Phi_n\!\!\cdot\!\Phi_1)
            \right]
            \otimes
            \bar{\mc{F}}_n$.
      \label{fig:pp2vv:1em}
    }
\end{figure}

\paragraph*{$\mathrm{pp}\to\nu_e\bar{\nu}_e$.}
As a simplified testing ground for the \QCDpEW \MCatNLO matching,
we turn to $\mathrm{pp}\rightarrow\nu_e\bar{\nu}_e$ production
in order to study a process without resonances in either
subtraction or parton evolution as the final state neutrinos
are uncharged under both \QCD and \QED and no dipoles span the
resonance.
We set the factorisation, renormalisation, and shower starting
scales to the neutrino-pair invariant mass,
$\mu_\mr{F}=\mu_\mr{R}=t_n=m_{\nu\nu}$.

In Fig.\ \ref{fig:pp2vv:1em}, we compare a fixed-order NLO \QCDpEW
calculation against \MCatNLO \QCDpEW truncated after
the first emission and split into $\mr{S}$- and $\mr{H}$-events.
We show the differential 0-to-1 jet resolution, $d_{01}$,
defined by clustering all partons
including photons with the $k_\mathrm{T}$-algorithm
\cite{Catani:1993hr} with $R=0.1$ using \FastJet \cite{Cacciari:2011ma},
and the transverse momentum of the neutrino pair as a proxy for
the $Z$ boson, $p_{\perp,Z}$.
For hard emissions and large $p_{\perp,Z}$, we observe that the
\MCatNLO result is dominated
by $\mr{H}$-events and closely reproduces the fixed order result.
For softer emissions, we are instead dominated by $\mr{S}$-events
which first exceed the fixed order result, by approximately 30\%
in the jet rate---this is the $\bar{\mr{B}}/\mr{B}$ prefactor to
the $\mr{S}$-event emissions kernel---then display the
characteristic Sudakov suppression as the scale of the emission,
and likewise $p_{\perp,Z}$, tends to zero.
The transition between $\mr{S}$- and $\mr{H}$-events occurs
at our chosen parton shower starting scale $t_n=m_{\nu\nu}$,
peaked around $m_Z$, limiting the resummation phase space.
We observe, however, that the $\mr{H}$-events induce sizeable
hard-emission corrections when $p_{\perp,Z}$ is as low
as 20\,GeV.

We further investigate the structure of our calculation by
splitting our generated events according to the five terms of
Eqs.\ \eqref{eq:methods:mcatnlo:Bbar} and \eqref{eq:methods:mcatnlo:H},
each associated with an underlying fixed-order contribution.
In particular, we split the $\mr{S}$-events into the Born
contribution $\mr{S}^{\mr{B}}$, the \QCD correction
$\mr{S}^\text{\QCD}$, and the \EW correction
$\mr{S}^\text{\EW}$ (see caption of Fig.\ \ref{fig:pp2vv:1em}).
Likewise, the $\mr{H}$-events are split into their \QCD and
\QED contributions $\mr{H}^\text{\QCD}$ and $\mr{H}^\text{\QED}$,
respectively, each containing only one type of real correction.

The $\mr{S}$-events are dominated by their Born contribution,
with their \QCD corrections providing a corrections of around
$+10\%$.
Their \EW corrections contribute just below $1\%$.
As all contributions share the same interleaved parton
shower and the structural simplicity of this process renders
the one-loop corrections nearly phase-space-independent,
$\mr{S}^{\mr{B}}$, $\mr{S}^\text{\QCD}$, and $\mr{S}^\text{\EW}$
share the same shape, with their normalisation difference
originating in the size of the loop corrections: about 10\%
and 1\% of the Born contribution, respectively.
The $\mr{H}$-events, on the other hand, cleanly separate
\QCD and \QED corrections and are dominated by \QCD
real emissions, while \QED emissions contribute only about
1\%.
This is unsurprising as, besides the larger coupling, the QCD
correction gives access to
gluon-induced processes and the corresponding large PDF effects.
We therefore recover the behaviour we expect from the first
\MCatNLO emission.

\begin{figure}[t!]
    \centering
    \includegraphics[width=0.49\textwidth]{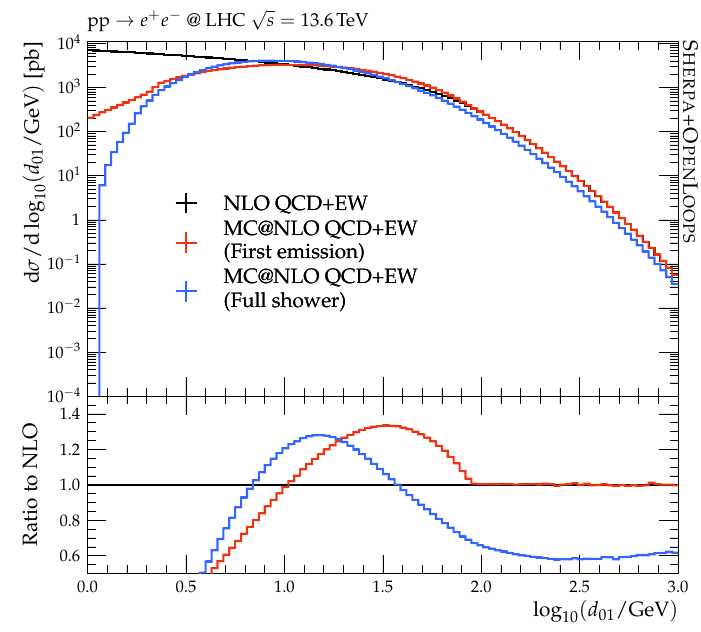}
    \includegraphics[width=0.49\textwidth]{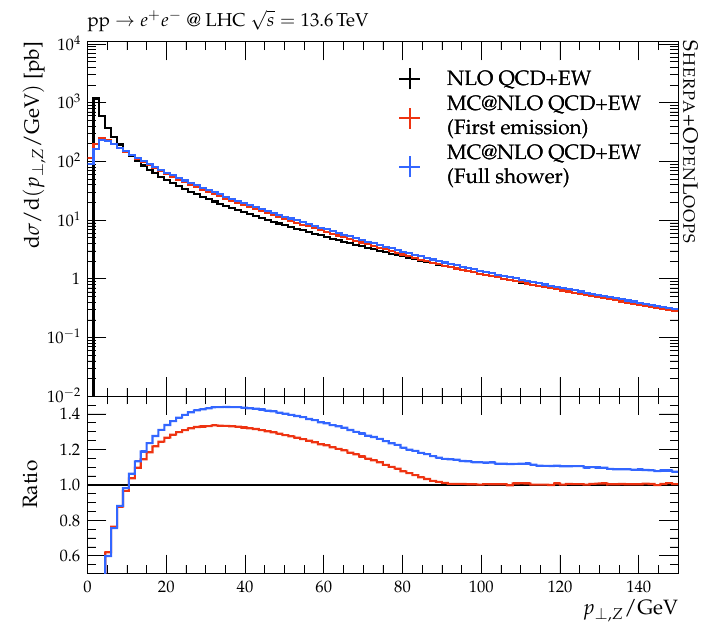}
    \caption{\label{fig:pp2vv:nlo_mcatnlo}Observables for $\mr{pp}\to \nu\bar{\nu}$ 
            at NLO \QCDpEW compared to a full \MCatNLO \QCDpEW calculation 
            as well as one truncated at the first emission.}
\end{figure}

Fig.\ \ref{fig:pp2vv:nlo_mcatnlo} shows the same fixed order
and truncated \MCatNLO results, now compared against the fully
showered \MCatNLO calculation.
It is imperative to note that all three $\mr{S}$-event contributions
effect the same interleaved \QCDpQED parton evolution, see
Sec.\ \ref{sec:methods:mcatnlo}, and each can produce \QCD and \QED
emissions.
As we would expect, including the full parton shower broadens the 
$p_{\perp,Z}$ distribution compared to the one-emission case
and gives a more pronounced Sudakov suppression in the jet rate.
The hardest jets are suppressed by approximately 40\% as the
hard partons have a higher probability to split later in the shower.

\begin{figure}[t!]
    \centering
    \includegraphics[width=0.49\textwidth]{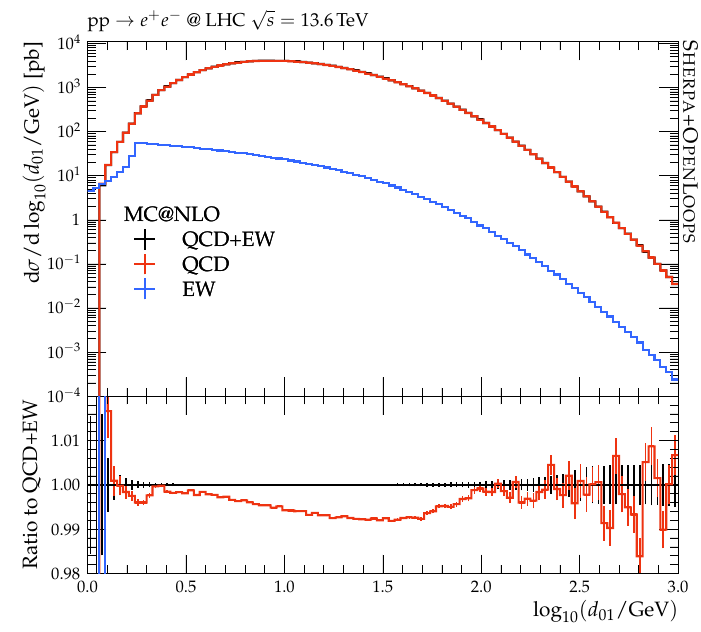}
    \includegraphics[width=0.49\textwidth]{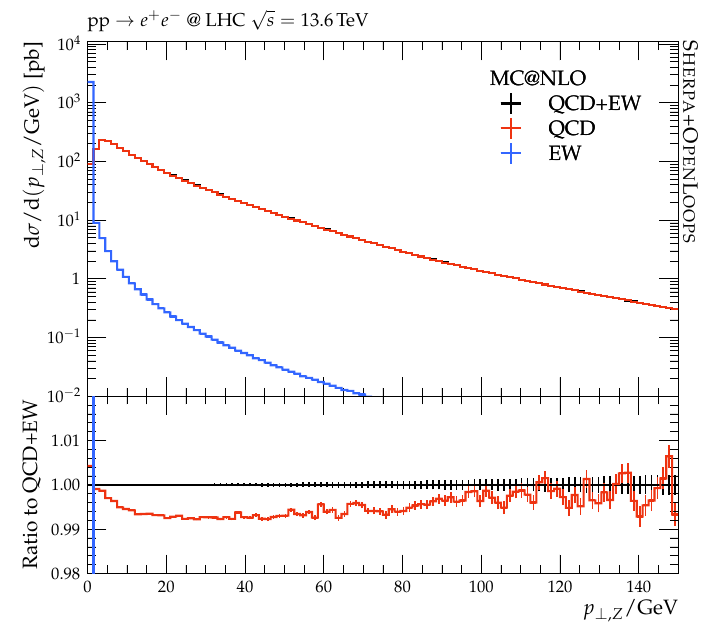}
    \caption{\label{fig:pp2vv:qcdew}Observables for $\mr{pp}\to \nu\bar{\nu}$ 
            calculated at \MCatNLO \QCDpEW, \QCD, and \EW.}
\end{figure}

Finally, we compare a full \MCatNLO \QCDpEW calculation
against separate \MCatNLO \QCD and \MCatNLO \EW results in
figure \ref{fig:pp2vv:qcdew} for the same observables.
As expected, \MCatNLO \QCDpEW closely aligns with the results
of \MCatNLO \QCD, with the addition of \EW matching appearing
as a sub-percent enhancement.
The parton shower infrared cutoff appears is now clearly visible
for \MCatNLO \EW as, due to the smaller coupling factor,
the Sudakov suppression of the first emission is much smaller
and, likewise, the Sudakov peak itself sits below the cutoff.

\paragraph*{$\mathrm{pp}\to e^+e^-$.}
We now move to consider a charged final state $\mathrm{pp}\to e^+e^-$
in the window around the $Z$ mass
in which \QED dipoles that span the $Z$ resonance can form.
Treating all leptons as massless in the hard scattering
calculation, we need to define their momentum in an infrared-safe
way: we dress all electrons and positrons in the
event by adding their four-momenta with those of all
photons in a cone of $R=0.1$ around the bare electron.
The leading dressed electron-positron pair is required
to have an invariant mass $60\,\GeV\leq m_{ee}\leq 120\,\GeV$,
a transverse momentum $p_{\perp,e^{\pm}}\geq 20\,\GeV$,
and rapidity $-2.5 \leq\eta_{e^{\pm}}\leq 2.5$.
We set all hard scales to the invariant mass of the dressed
lepton pair, $\mu_\mr{F}=\mu_\mr{R}=t_n=m_{ee}$.
In addition to our standard \MCatNLO \QCDpEW calculation
we also include a \MCatNLO calculation where we include
the modified $\mr{H}$-events
of Eq.\ \eqref{eq:methods:mcatnlo:Hsud}, dubbed $\mr{H}^\text{sud}$,
that applies an additional Sudakov factor for the evolution
between the scales $t_n$ and $t_{n+1}$.

\begin{figure}[t!]
    \centering
    \includegraphics[width=0.49\textwidth]{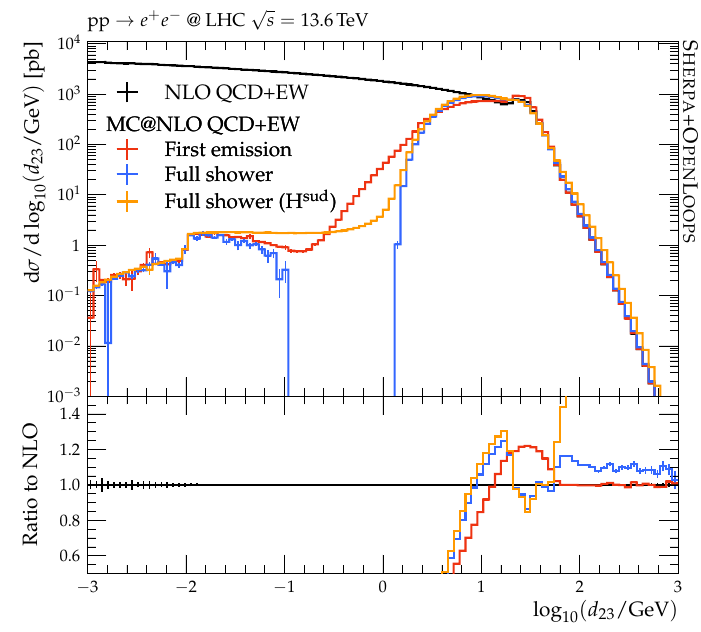}
    \hfill
    \includegraphics[width=0.49\textwidth]{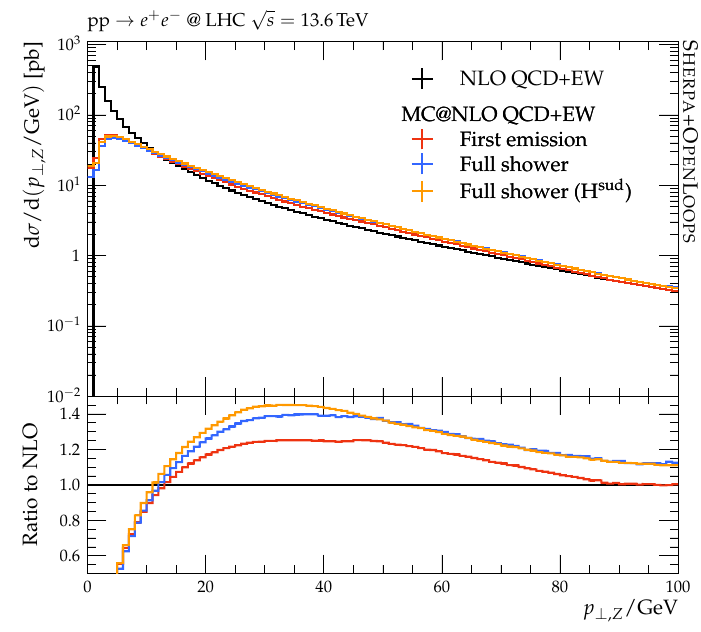}\\
    \includegraphics[width=0.49\textwidth]{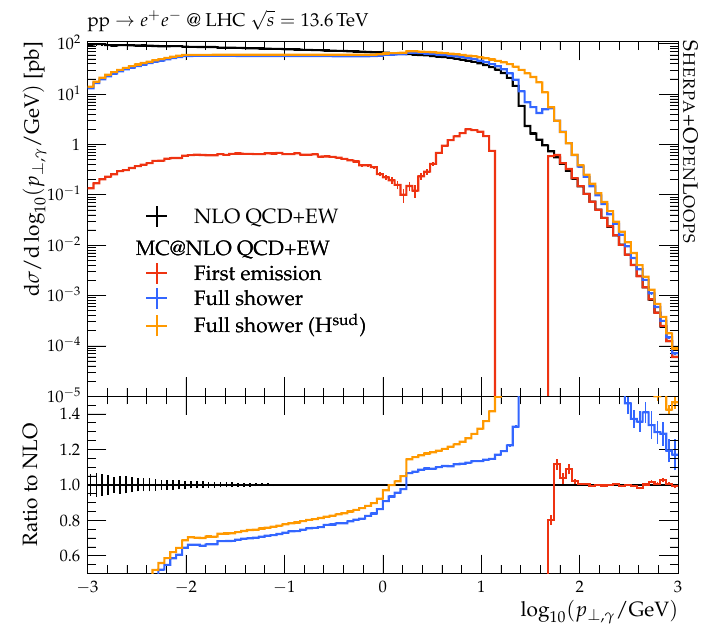}
    \hfill
    \includegraphics[width=0.49\textwidth]{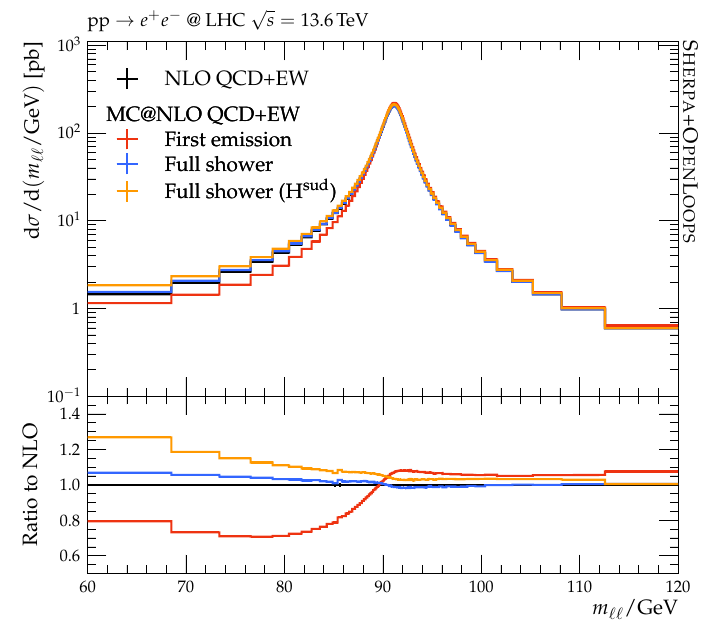}
    \caption{
      Observables for $\mr{pp}\to e^+e^-$
      at NLO \QCDpEW compared to a full \MCatNLO \QCDpEW calculation,
      with and without a Sudakov applied to the $\mr{H}$-events,
      as well as one truncated at a the first emission.
      \label{fig:pp2ee:nlo_mcatnlo}
    }
\end{figure}

As for the neutrino case, we start by comparing our
\MCatNLO \QCDpEW calculations, both truncated and with
the full parton shower, against the fixed-order
NLO \QCDpEW calculation in Fig.\ \ref{fig:pp2ee:nlo_mcatnlo}.
To characterise the emission pattern of our interleaved
parton shower, we include the bare partonic leptons
and photons in the jet finding algorithm.
Consequently, the first emission
is now predominantly described by the $d_{23}$ jet resolution
rather than $d_{01}$.
As before, we observe that the fixed-order NLO calculation
and truncated \MCatNLO  agree closely for hard emissions
in $d_{23}$.
At lower resolution scales the effects of first the
$\bar{\mr{B}}/\mr{B}$ enhancement and then the combined
Sudakov factor dominate the spectrum until we reach
the infrared cutoffs of the parton shower, $t_c=1\,\GeV^2$
for final state \QCD evolution and $3\,\GeV^2$ for initial
state \QCD and \QED evolution.\footnote{%
  Below this scale the comparison to the fixed-order
  calculation is no longer meaningful as the latter's infrared
  radiation pattern is not regulated and will continue
  into the well-known singularity at $d_{23}\to 0$.
}
As the \QCD radiation which dominates while it is active
drops away to leave only final state \QED radiation
until its own cutoff $t_c=m_e^2$,
which are suppressed by almost three orders of magnitude
due to the extremely small Sudakov factor.
Please note, we have extended the reach of the observable
into this far-infrared region compared to the
$\mr{pp}\to\nu_e\bar{\nu}_e$ case to characterise the
\QED radiation off the final state leptons.
This drop is made sharper by including the full shower,
and without the additional Sudakov factor in the
$\mr{H}$-events of the \MCatNLOsud the differential
cross section turns negative immediately below this cutoff,
where we have only relatively soft photon emissions.
This is an unphysical artefact of uncontrolled higher order
corrections, which we see is removed by including the Sudakov factor
in $\mr{H}^\text{sud}$, see Sec.\ \ref{sec:methods:mcatnlo}.

We find similar effects in the $p_{\perp,Z}$ spectrum,
now defined as the transverse momentum of the dressed
electron-positron pair.
Again, the inclusion of the additional Sudakov in \MCatNLOsud
cures the unphysical behaviour of some of the
$\mr{H}$-events. This modification
removes the small kink in the transverse
momentum distribution of the reconstructed $Z$ boson
spectrum at $p_{\perp,Z}\approx\tfrac{1}{2}\,m_Z$ that
appears without it.

We turn now to the inclusive photon transverse momentum
$p_{\perp,\gamma}$, where each photon that has \emph{not}
been recombined into a dressed lepton is entered
seperately into the histogram, shown in the lower left
plot of Fig.\ \ref{fig:pp2ee:nlo_mcatnlo}.
While its integral, like that the inclusive jet rate in
jet production, is not the production cross section of the
process, it does, however, allow to study the distribution of
photons that are not collinear with our final-state leptons.
As a non-inclusive observable, all the calculations shown
only have LO accuracy.
Comparing first the truncated \MCatNLO calculation to the
fixed-order calculation, we observe that, while the hard photon
spectrum described through the $\mr{H}^\text{\QED}$-events
matches the fixed order one, the impact of the large \QCD
Sudakov factor that the interleaved parton shower evolution
invariably applies to the \QED emission suppresses the spectrum
by more than an order of magnitude wrt.\ the fixed-order
description.
Worse, the $p_{\perp,\gamma}$ spectrum turns negative for
$15\,\GeV\lesssim p_{\perp,\gamma}\lesssim 40\,\GeV$.
And while including the full parton shower evolution restores
the rate to roughly those predicted at fixed-order, it exemplifies
that the source of most photons in this region are not the
primary emissions
of the core process, but rather emission that are displaced by
harder \QCD splittings and occur later in the parton evolution.
While this may seem to degrade the formal accuracy of our
calculation in this soft-photon regime, it highlights the
importance of interleaved parton evolution and the inadequacy
of the fixed-order calculation in this regime.
Nonetheless, these artifacts
survive as cusps in the unmodified \MCatNLO \QCDpEW calculation.
The inclusion of the additional Sudakov factor in the
$\mr{H}$-events in our \MCatNLOsud
cures this behaviour, and moves the drop-off in the spectrum to
the expected value of $p_{\perp,\gamma}\approx\tfrac{1}{2}\,m_Z$.

Discussing the invariant mass spectrum of the
dressed electron-positron pair in the lower right panel of
Fig.\ \ref{fig:pp2ee:nlo_mcatnlo},
this observable is, too, largely driven by real photon
radiation.
Here, examining first again the truncated \MCatNLO calculation,
we observe the suppression of real photon radiation by the large
\QCD Sudakov factor, leading to an underestimate of the spectrum
below the peak.
The full \MCatNLO calculation including the complete subsequent
parton evolution relaxes the first-emission constraint and seems
to lead to a much
better agreement, but the positive $\order(\alpha^2)$
(and higher) corrections of the full parton evolution hide a
residual shortfall.
With the inclusion of the additional $\mr{H}$-event Sudakov factor
in the \MCatNLOsud the remaining discrepancy is remedied and good
agreement, accounting for the additional higher-order
\QED corrections, is found.

\begin{figure}[t!]
    \centering
    \includegraphics[width=0.49\textwidth]{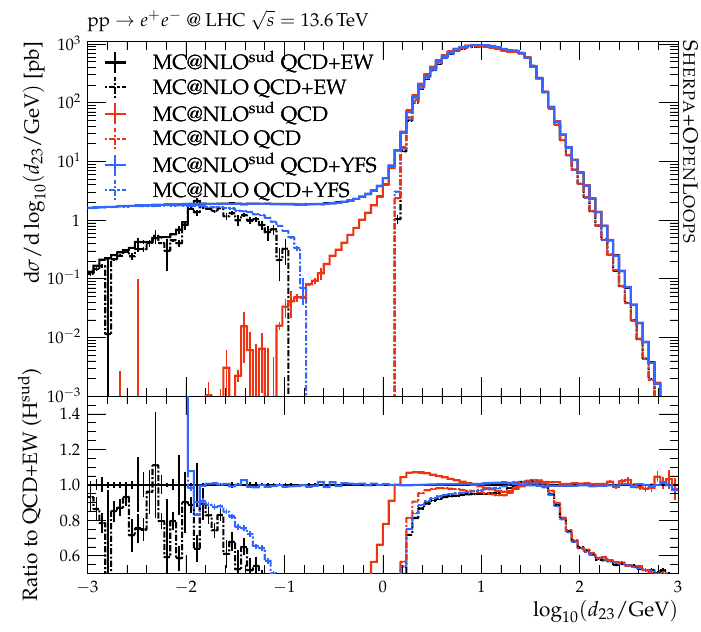}
    \hfill
    \includegraphics[width=0.49\textwidth]{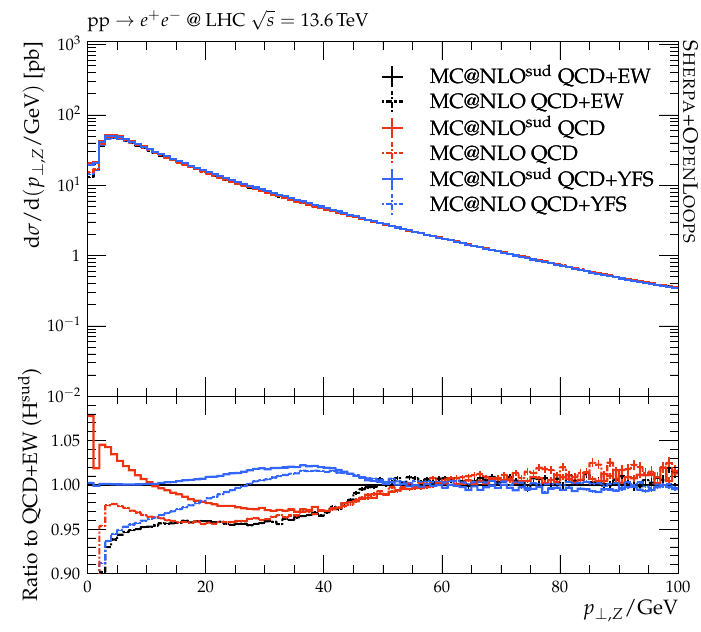}\\
    \includegraphics[width=0.49\textwidth]{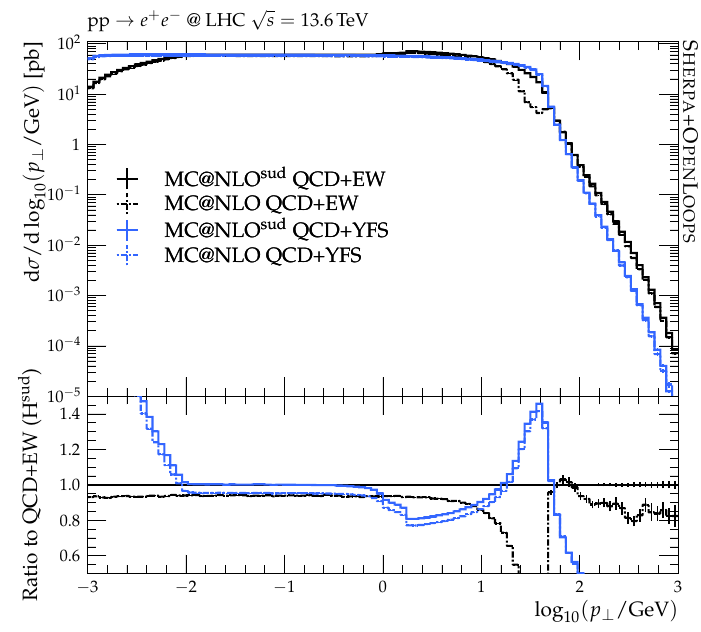}
    \hfill
    \includegraphics[width=0.49\textwidth]{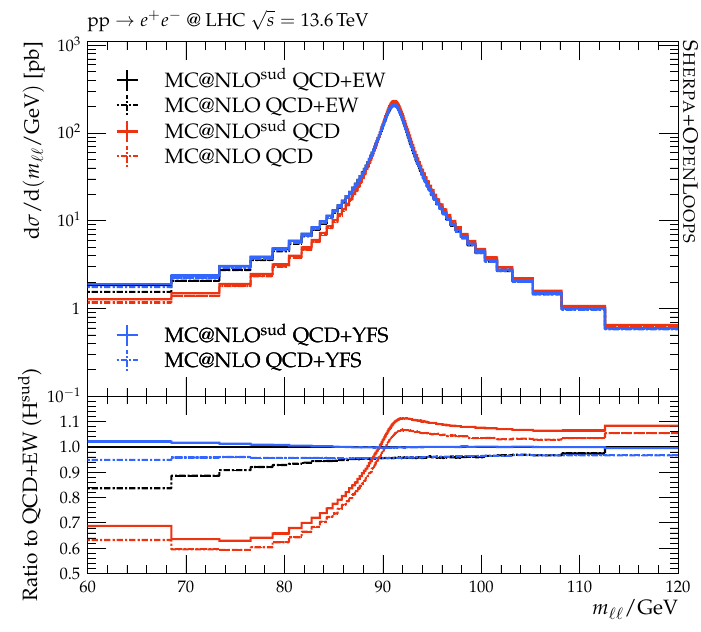}
    \caption{
      Observables for $\mr{pp}\to e^+e^-$
      calculated at \MCatNLO \QCDpEW, \MCatNLO \QCD, and \MCatNLO \QCDpYFS,
      along with the same calculations using \MCatNLOsud.
      \label{fig:pp2ee:qcdew}
    }
\end{figure}

Lastly, in Fig.\ \ref{fig:pp2ee:qcdew} we compare the
\MCatNLO \QCDpEW and the \MCatNLOsud \QCDpEW calculations,
against the existing NLO-parton-shower-matching
in \QCD only, with \QED corrections to the $Z\to e^+e^-$
decay generated through the soft-photon resummation in the
Yennie-Frautschi-Suura (\YFS) scheme \cite{Yennie:1961ad,Schonherr:2008av},
dubbed \MCatNLO \QCDpYFS and \MCatNLOsud \QCDpYFS.
This soft-photon resummation provides an exclusive but unordered
description of \QED corrections to the decay subprocess,
including hard-emission corrections up to NNLO \QED+NLO \EW accuracy
\cite{Schonherr:2008av,Krauss:2018djz}.
An \MCatNLO QCD calculation\xspace
, with and without the additional Sudakov in the $\mr{H}$-events,
without any \QED or \EW higher-order
effects is added to gauge the size of the \QED/\EW corrections
and their deviation between the different approaches.

In the jet rate, in the top left panel, we notice that the hard
emissions are dominated by the \QCD dynamics of each calculation
and agree on the level of a few percent with their \QCD-only
calculation.
\QED splittings are relevant at lower $d_{23}$, and the addition
of \QED initial state radiation improves the spectrum in the
region of approximately $d_{23}\in[5,\tfrac{1}{2}\,m_Z]\,\GeV$,
though this correction is much smaller with \MCatNLOsud.
Below that scale, final state radiation off the leptons dominates
the \QED corrections.
Interference effects only included in the full
\MCatNLO{}$^\text{(sud)}$ \QCDpEW calculation are small.
Below the \QCD and initial state \QED parton shower cutoffs,
we can examine the final state \QED radiation pattern in isolation.
After curing the unphysical features induced by the large
uncancelled interleaved Sudakov factor in the \MCatNLOsud-type
calculations, the final state radiation pattern
agrees very well with the established \YFS result.

Turning to considering $p_{\perp,Z}$ in the top right panel
of Fig.\ \ref{fig:pp2ee:qcdew}, we observe that, similarly,
the high-$p_{\perp,Z}$ region is dominated by \QCD effects
as these are the hardest emissions from the initial state,
maximising recoil against the electron pair.
A small impact originating in the description of final-state
radiation off the leptons, lowering their combined momentum
on average even after dressing, is visible.
The impact of additional interleaved Sudakov factor can be
seen in the region of $p_{\perp,Z}\lesssim \tfrac{1}{2}\,m_Z$,
where through reducing the negativity of the $\mr{H}$-events
in this process, the differential cross section is increased
by about 5\%.
The \EW corrections, driven again by \QED radiative effects,
are largest at moderate $p_{\perp,Z}\approx\tfrac{1}{2}\,m_Z$.
The comparison of the full \QCDpEW results with the final-state-only
\YFS counterparts reveals an intricate interplay between
ISR and FSR, increasing and reducing the differential cross
section, respectively, until FSR dominates below
$p_{\perp,Z}\lesssim 10\,\GeV$.

On the other hand, when we consider the $p_{\perp,\gamma}$
distribution in the bottom left panel of
Fig.\ \ref{fig:pp2ee:qcdew},
we instead find that the interleaved \QCDpQED splittings
leads an enhancement, especially to hard photons,
beyond the flat factor provided by the larger total
cross section.
This is because the \QCD shower adds more quarks to the
event which are able to partake in both \QCD and \QED
splittings, so there are more potential photon emitters.

The invariant mass distribution of the dressed
electron-positron pair is shown in the bottom right panel
of Fig.\ \ref{fig:pp2ee:qcdew}.
As expected, \QCD alone gives a poor description
of $m_{ee}$ as \EW corrections, in particular
\QED radiation off the final state leptons, are of $\order(1)$.
Adding in the \YFS resummation captures most of this and gives
approximately the same shape as the \MCatNLOsud \QCDpEW when
we include the additional Sudakov form factor.
Without it, however, the \MCatNLO \QCDpEW does not
reproduce this distribution below the $Z$ peak, due to its
large higher order artefacts even in inclusive observables.

\begin{figure}[t!]
    \centering
    \includegraphics[width=0.49\textwidth]{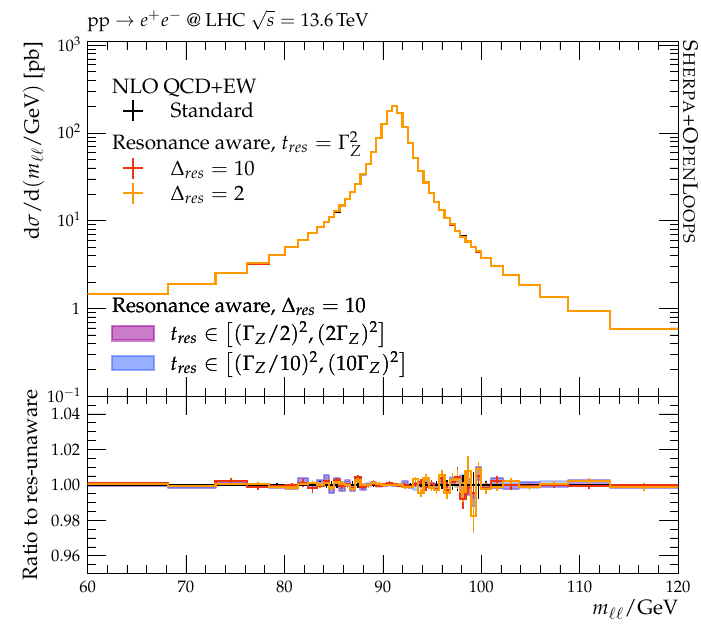}
    \hfill
    \includegraphics[width=0.49\textwidth]{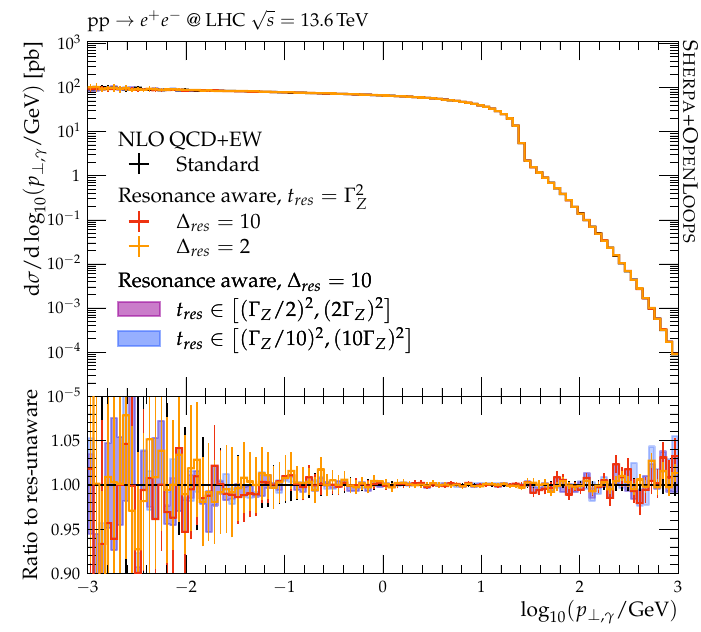}
    \caption{
      Observables for $\mr{pp}\to e^+e^-$
      calculated at NLO \QCDpEW using standard Catani-Seymour
      dipole subtraction (black) and using resonance-aware dipoles
      (red and orange).
      We show variations of both $\tres$ and $\Deltares$.
      \label{fig:pp2ee:mcfo}
    }
\end{figure}

\paragraph*{Resonance-aware subtraction of $\mr{pp}\to e^+e^-$.}
Having demonstrated the validity of the interleaved \MCatNLO, 
we now consider our resonance-aware subtraction scheme. 
We first test this at fixed order to verify to that
the results are independent of the subtraction scheme
and that there are no unsubtracted singularities,
which would otherwise spoil convergence.
Fig.\ \ref{fig:pp2ee:mcfo} shows an NLO \QCDpEW calculation
using both the standard dipoles $\mr{D}_{ij,k}^a$
and the resonance-aware dipoles $\mr{D}_{ij,k}^{a,\text{res}}$.
For the latter, we also vary the resonance parameters,
showing $\Deltares=2,\,10$ for $\tres=\Gamma_Z^2$ and
the regions spanned by varying around $\sqrt{\tres}=\Gamma_Z$
by factors of 2 and 10 at $\Deltares=10$.
In all cases, we find that the results differ only by statistical
variations, showing that, indeed, the fixed order result
is independent of our choice of subtraction scheme.
We expect numerical stability to grow slightly worse if $\tres$ is
too small because we do not perform the exact subtraction until
we are very close to the singularities, leaving the possibility
of numerical artefacts, however the results always converge well,
demonstrating that we still capture all singular limits.
Our resonance-aware treatment therefore satisfies the necessary
requirements of a fixed order subtraction scheme.

\begin{figure}[t!]
  \includegraphics[width=0.49\textwidth]{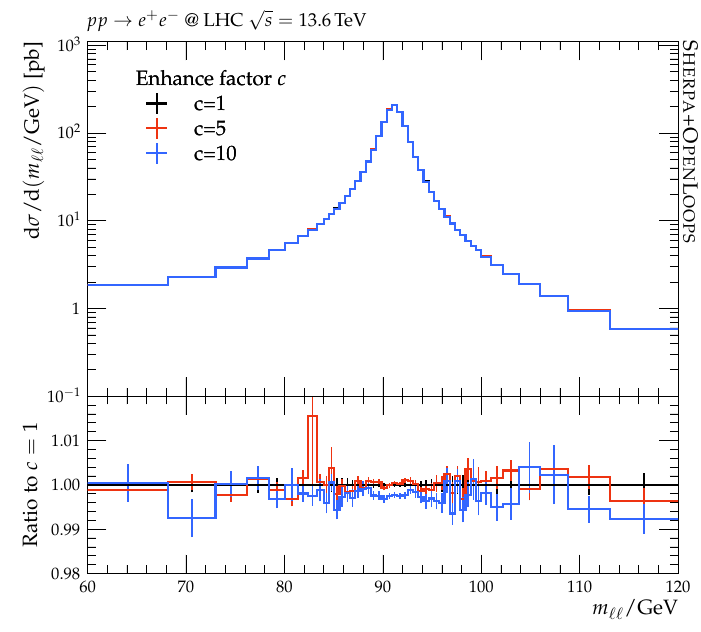}
  \hfill
  \includegraphics[width=0.49\textwidth]{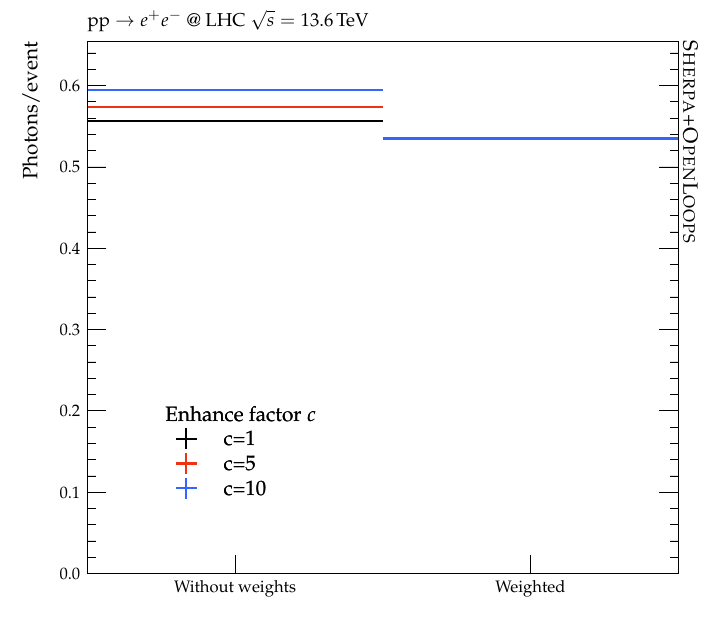}\\
  \includegraphics[width=0.49\textwidth]{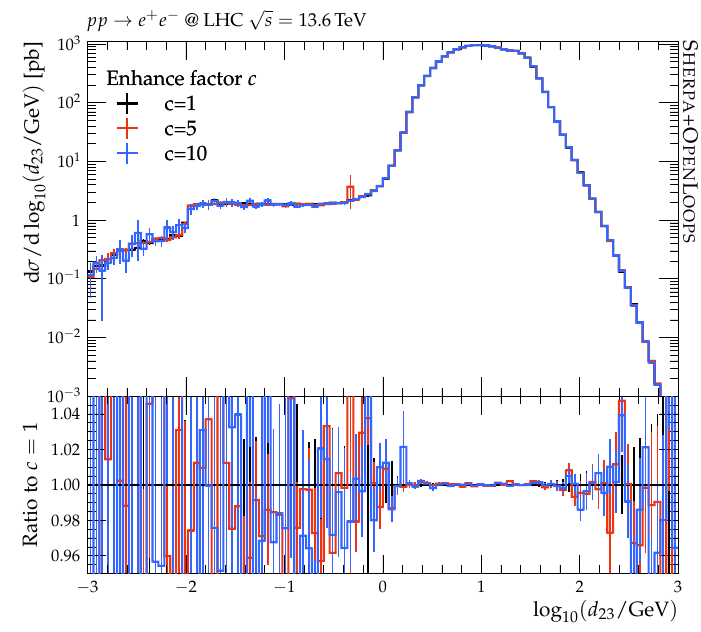}
  \hfill
  \includegraphics[width=0.49\textwidth]{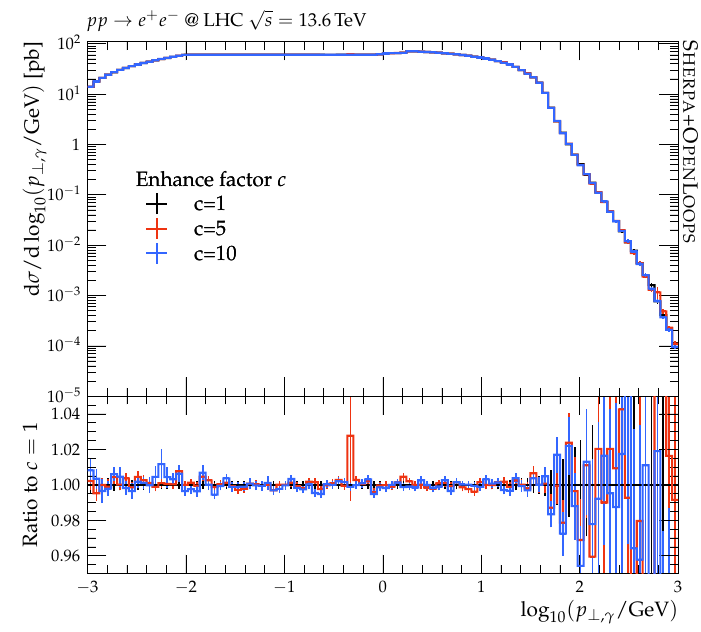}
  \caption{
    Observables for resonance aware $\mr{pp}\to e^+e^-$ with
    \MCatNLO \QCDpEW with enhanced \QED splittings in the
    first emission.
    We also show the average number of photons generated per event,
    with and without accounting for event weights.
    \label{fig:pp2ee:efacs}
  }
\end{figure}

\paragraph*{Statistically enhancing \QED emissions.}
Finally, we demonstrate the effects of enhanced \QED splittings
in Fig.\ \ref{fig:pp2ee:efacs} using the resonance-aware
\QCDpEW \MCatNLO.
\QED emissions are enhanced by replacing
$\frac{\mr{D}_n}{\mr{B}_n} \to c\,\frac{\mr{D}_n}{\mr{B}_n}$
in the generation of \MCatNLO $\mr{S}$-events,
for some constant enhance factor $c$,
then correcting for this in the event weight,
see Sec.\ \ref{sec:methods:mcatnlo}.
No enhancement to the subsequent parton shower is applied, as it is
most important to improve photon sampling in the first emission.
We show results for $c=5$ and 10 compared against an unenhanced sample
of the same number of events.
The average number of photons generated within the
infrared cutoffs of Sec.\ \ref{sec:methods:ps}
per event, ignoring event weights (shown in the left column),
increases for larger $c$ as desired, but we see that the weights
exactly compensate for this effect (shown in the right column),
with excellent agreement between all three samples.

The unenhanced dilepton mass distribution in the top left panel of
Fig.\ \ref{fig:pp2ee:efacs} is exactly reproduced when using
an enhance factor of $c=5$.
However, we see a flat, sub-percent decrease in the differential
cross section when increasing to $c=10$.
This effect is not seen in radiative observables, such as
$d_{23}$ and $p_{\perp,\gamma}$ shown in the bottom row of
Fig.\ \ref{fig:pp2ee:efacs}, and is an
artefact of the exceeding rarity of non-radiative events to
appropriately correct their contribution to the sample.
Larger enhance factors also lead to a larger spread in event
weights, which degrades the statistical performance of the sample,
as indicated by the statistical uncertainties shown.
For this reason, there is a limit by how much the \QED splittings
can be enhanced before this degradation becomes prohibitive.
While a \QED enhance factor of 10 may seem reasonable, since it
makes $\alpha$ of comparable size to $\alphaS$, we find that it
is already too large, introducing unphysical artefacts on the
level of 0.5\%.
An enhancement with $c=5$, however, while not producing any
artefacts, does not appreciably decrease the statistical
uncertainty either.
We thus choose not to artificially enhance the \QED splittings
in the matched first emission of our parton evolution in the
following.

\subsection{Resonance-aware parton-showered-matched predictions}
\label{sec:results:prediction}

Having validated both the interleaved \MCatNLOsud \QCDpEW implementation
and the resonance aware subtraction scheme,
we now use these in tandem to make predictions for observations of
$\mathrm{pp}\to e^+e^-$ at the LHC and quantify the extent to which
the resonance is distorted by the standard treatment.
We use the same cuts as in Sec.\ \ref{sec:results:validation}.

\begin{figure}[t!]
    \centering
    \includegraphics[width=0.49\textwidth]{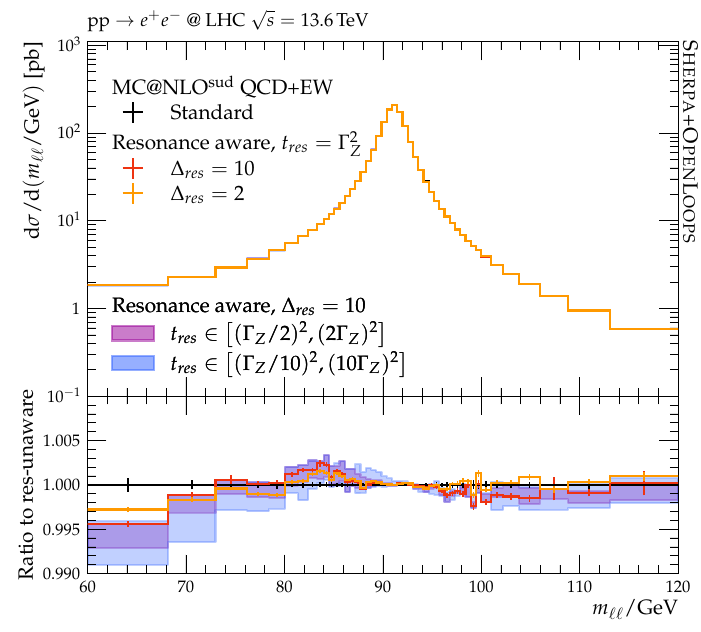}
    \hfill
    \includegraphics[width=0.49\textwidth]{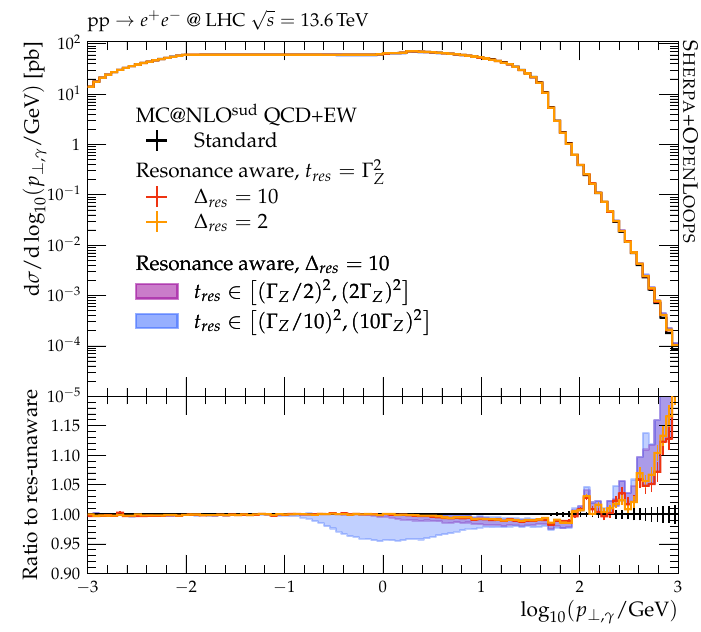}
    \caption{
      Observables for $\mr{pp}\to e^+e^-$
      calculated at \MCatNLOsud \QCDpEW
      standard dipoles and with resonance-aware dipoles.
      We show variations of both $\tres$ and $\Deltares$.
      \label{fig:pp2ee:resaware}
    }
\end{figure}

In Fig.\ \ref{fig:pp2ee:resaware}, we present a resonance-aware
\MCatNLOsud \QCDpEW calculation of the dressed dilepton mass $m_{ee}$
and the photon transverse momentum $p_{\perp,\gamma}$ for
electron-positron pair production at the LHC.
As in Fig.\ \ref{fig:pp2ee:mcfo}, we compare against the same
calculation with resonance-unaware parton shower evolution
as well as showing the
effects of varying the resonance-awareness parameters
$\tres$ and $\Deltares$, see Sec.\ \ref{sec:methods:resaware}.
Examining the invariant mass distribution in its right panel,
we observe that all calculations agree closely at the $Z$
peak itself, independent of the resonance-awareness parameters.
In this region, all dipole parameter settings identify the
configuration as resonant and prepare the subtraction and
parton evolution in the same way.
Likewise, as discussed in Sec.\ \ref{sec:methods:resaware},
when the resonance is on-shell, the sum of same-sign and
opposite-sign resonance-spanning dipoles vanishes for
neutral resonances.
For invariant masses slightly below the $Z$ peak,
there is a 0.2\% change to the shape,
growing to 0.5\% at $m_{ee}\approx 60\,\GeV$.
For higher invariant masses beyond the $Z$ peak
the resonance-aware and -unaware results largely coincide,
with differences depending on the precise definition of
what constitutes a resonance, $\Deltares$, amounting to
less than 0.2\%.
Naturally, reducing $\Deltares$ brings the result closer to the 
resonance-unaware case because fewer events are identified as resonant;
however, there is still a clear change in the shape of both distributions
even for this very conservative definition of a resonant event.
Similarly, we see the same behaviour for all values of $\tres$.

Turning to the photon transverse momentum in the right panel of
Fig.\ \ref{fig:pp2ee:resaware}, we observe that while photon emissions
within $p_{\perp,\gamma}\in[1,100]\,\GeV$ are reduced
in the resonance-aware calculation on the level of up
to 2\%, the hardest photon emissions are enhanced by up to 15\%
at $p_{\perp,\gamma}\approx 1\,\TeV$, mainly through the missing
suppression through ISR-FSR interference effects induced
through the resonance-spanning IF and FI dipoles of opposing sign beyond NLO.
The factor-10 uncertainty band becomes very large close to
$p_{\perp,\gamma}=1\,\GeV$, primarily due to the downward
variation to $\tres=(\Gamma_Z/10)^2$, however it should be
noted that, as discussed at the end
of the previous section, the results are not expected to be
entirely reliable with very low $\tres$, because they depend
on the approximate cancellation of the neglected dipoles until
very close to the singular limits.
We see, therefore, that there is indeed a sub-percent distortion
of the resonance in the standard calculation of $\mr{pp}\to e^+e^-$
at \MCatNLOsud \QCDpEW and that a resonance-aware treatment is necessary
to accurately obtain high precision predictions.

\section{Conclusions}
\label{sec:conclusions}

In this paper we have presented the first automated matching of
NLO \QCDpEW to an interleaved \QCDpQED parton shower
using the \MCatNLO matching method in the Catani-Seymour
dipole formalism.
Furthermore, we have extended this method to include the treatment
of colour- and charge-neutral resonances by suitably constraining dipoles
that span the resonance within a region of phase space defined
by the ability of an emission to resolve the resonance and
the virtuality of the resonant propagator.
We have presented the predictions of this resonance-aware \MCatNLO \QCDpEW
for Drell-Yan lepton pair production at hadron colliders.

We have carefully validated the \MCatNLO \QCDpEW against fixed-order 
calculations and against the existing \MCatNLO \QCD implementation in \Sherpa,
combined with \QED final-state radiative corrections in the YFS formalism 
\cite{Yennie:1961ad,Schonherr:2008av}.
We showed that unphysical artefacts appear in the standard 
formulation of \MCatNLO when used for interleaved matching
due to the effects of uncontrolled $O(\alphaS\alpha)$ terms
which are left behind in the modified subtraction of parton
shower emission kernels.
We have demonstrated that these artefacts can be removed by 
introducing an additional physically motivated Sudakov factor
to the $\mr{H}$-events to suppress these higher order terms.
The cost of this modification is that the inclusive cross section, while
retaining its accuracy to NLO, is modified by higher order effects.

We have illustrated the predictions of our interleaved \MCatNLO \QCDpEW 
matching, and in particular the resonance-aware modifications,
using lepton pair production, $\mr{pp}\to e^+e^-$.
Here we see that the resonance-aware treatment corrects a distortion 
of the dilepton mass distribution of up to a few per mille, along with 
a percent level change in the transverse momentum of hard photons, 
with the effect becoming larger for the hardest tail.
We have provided uncertainty estimates of our predictions by varying 
the parameters we use to define the kinematic region in which our 
resonance-aware modifications take effect.

While our standard interleaved \MCatNLO \QCDpEW is fully automated,
our implementation of resonance-aware matching is currently
process-specific, however the algorithm is entirely general
and could easily be automated if combined with
automated resonance identification \cite{Sherpa:2024mfk,Kallweit:2017khh}.
For (colour-)charged resonances, however, the treatment presented here is
insufficient, because the production-decay factorisation
necessitates the introduction of
new splitting functions in which the (colour-)charged
resonant particle may act as both emitter and spectator.
These extensions will be the subject of future work.

\subsection*{Acknowledgements}

JR and MS are funded by the Royal Society through a University Research Fellowship
(URF\textbackslash{}R1\textbackslash{}180549, URF\textbackslash{}R\textbackslash{}231031) and Enhancement Awards
(RF\textbackslash{}ERE\textbackslash{}210397,
 RGF\textbackslash{}EA\textbackslash{}181033 and
 CEC19\textbackslash{}100349).
MS further acknowledges funding from the STFC Grants
No.\ ST/T001011/1 and ST/P006744/1.
LF is supported by the Leverhulme Trust under grant LIP-2021-014.
LF also acknowledges support from an STFC studentship 
under the STFC training grant ST/P001246/1 for the early parts of this work.

\bibliographystyle{amsunsrt_modpp}
\bibliography{journal}
\end{document}